\documentclass[a4paper,12pt]{article}
\usepackage[top=3cm,bottom=3.cm,left=2.5cm,right=2.5cm]{geometry}
\usepackage{graphicx}
\usepackage{amsmath, amsthm, amsfonts,multirow,float,moreverb}
\usepackage{authblk}
\usepackage[flushleft]{threeparttable}
\usepackage{booktabs}
\usepackage[titletoc,title]{appendix}
\usepackage{etoolbox}
\patchcmd{\appendices}{\quad}{: }{}{}
\usepackage{setspace}
\usepackage{rotating}
\usepackage[numbers]{natbib}
\usepackage[pdftex]{hyperref}
\title{Missing binary outcomes under covariate dependent missingness in cluster randomised trials}
\author[1]{Anower Hossain}
\author[1]{Karla Diaz Ordaz}
\author[2]{Jonathan W. Bartlett  }
\affil[1]{\small Department of Medical Statistics\\ London School of Hygiene and Tropical Medicine}
\affil[2]{\small Statistical Innovation Group, AstraZeneca}
\date{}
\begin{document}
\maketitle
\newcommand{\add }{\added}
\newcommand{\del }{\deleted}
\newcommand{\rep }{\replaced}
\begin{abstract}
	Missing outcomes are a commonly occurring problem for cluster randomised trials, which can lead to biased and inefficient inference if ignored or handled inappropriately. Two approaches for analysing such trials are cluster-level analysis and individual-level analysis. In this study, we  assessed the performance of unadjusted cluster-level analysis, baseline covariate adjusted cluster-level analysis, random effects logistic regression (RELR) and generalised estimating equations (GEE) when  binary outcomes are missing under a baseline covariate dependent missingness mechanism. Missing outcomes were handled using complete records analysis (CRA) and multilevel multiple imputation (MMI). We analytically show that cluster-level analyses for estimating risk ratio (RR) using complete records are valid if the true data generating model has log link and the intervention groups have the same missingness mechanism and the same covariate effect in the outcome model.  We performed a simulation study considering four different scenarios, depending on whether the missingness mechanisms are the same or different between the intervention groups and whether there is an interaction between intervention group and baseline covariate in the outcome model. Based on the simulation study and analytical results, we give guidance on the conditions under which each approach is valid.
\end{abstract}
\maketitle
\footnotetext[2]{Email: anower.hossain@lshtm.ac.uk}
\section{Introduction}
\label{intro}
Cluster randomised trials (CRTs), also known as group randomised trials, are increasingly being used to evaluate the effectiveness of interventions in health services research \cite{campbell2007developments, Donnerandklar2000}. The unit of randomisation for such trials are identifiable clusters of individuals such as medical practices, schools, or entire communities. However, individual-level outcomes of interest are observed within each cluster.  One important feature of CRTs is that the outcomes of individuals within the same cluster are more likely to be similar to each other than those from different clusters,  which is usually quantified by the intraclass correlation coefficient (ICC, denoted as $ \rho $). Although typically in primary care and health research the value of ICC is small ($ 0.001 <\rho<0.05 $) \cite{murray_blitstein2003}, it can lead to substantial variance inflation factors and should not be ignored \cite{Donnerandklar2000, murray1998}. This is because ignoring the dependence of the outcomes of individuals within the clusters will underestimate the variance of the intervention effect estimates and consequently give inflated Type I error rates \cite{Murray2004}. It is well known that the power and precision of CRTs are lower compared to trials that individually randomise the same number of units \cite{Donnerandklar2000}. However, in practice, CRTs have several advantages including that the nature of the intervention itself may dictate its application at the cluster level, less risk of intervention contamination and administrative convenience \cite{hayes2009}. These advantages are sometimes judged by researchers to outweigh the potential loss of statistical power and precision.

Missing data are a commonly occurring threat to the validity and efficiency of CRTs. In a systematic review of CRTs published in English in 2011, 72\% of trials had missing values either in outcomes or  in covariates or in both, and only 34\% of them reported how missing data had been handled \cite{DiazOrdaz2014}. Dealing with missing data in CRTs is complicated because of the clustering of the data.  In statistical analysis, if there are missing values, an assumption must be made about the relationship between the probability of data being missing and the underlying values of the variables involved in the analysis. The mechanisms which caused the data to be missing can be classified into three broad categories. Data are missing completely at random (MCAR) if the probability of missingness is independent of the observed and unobserved data. MCAR is generally a very  restrictive assumption and is unlikely to hold in many studies. A more plausible assumption is missing at random (MAR) where, conditioning on the observed data, the probability of missingness is independent of the unobserved data. Missing not at random (MNAR) is the situation where the probability of missingness depends on both the observed and unobserved data. In CRTs, an assumption regarding missing outcomes that is sometimes plausible is that missingness depends on baseline covariates, but conditioning on these baseline covariates, not on the outcome itself. We refer to this as covariate dependent missingness (CDM). This is an example of MAR when baseline covariates are fully observed.  In this paper, we will consider the case of a binary outcome which is partially observed, and assume that all baseline covariates are fully observed.

Two approaches for analysing CRTs are cluster-level analyses, which derive summary statistics for each cluster, and individual-level analyses, which use the data for each individual in each cluster \cite{hayes2009}. Complete records analysis (CRA) and multiple imputation (MI) (described in Section \ref{Methods_handling_missing_data}) are the most commonly used methods for handling missing data. A number of recent studies have investigated how to handle missing binary outcomes in CRTs under the assumption of CDM \cite{Ma2011, Ma2012comparing, Ma2013, Caille2014}. However, as we describe in detail in Section \ref{Methods_handling_missing_data}, these previous studies simulated datasets in ways which arguably do not correspond to how data arise in CRTs raising doubt about their conclusions.   

In the case of missing outcome under MAR for individually randomised trials, Groenwold \textit{et al.} \cite{Groenwold2012} showed that CRA with covariate adjustment and MI give similar estimates as long as the same set of predictors of missingness are used. It can be anticipated that a similar result holds for CRTs. In the case of missing continuous outcomes in CRTs, Hossain \textit{et al.} \cite{Hossain2016} showed that there is no gain in terms of bias or efficiency of the estimates using MI over CRA adjusted for covariates, where both approaches used the same set of baseline covariates and modelling assumptions. Therefore in situations where they are equivalent, CRA is clearly preferable. 

All of these previous studies \cite{Ma2011, Ma2012comparing, Ma2013, Caille2014} considered only individual-level analysis and estimated odds ratio (OR)  as a measure of intervention effect. The risk difference (RD) or risk ratio (RR) may be of interest as measures of intervention effect, and have a number of advantages over OR \cite{Davies1998}. For example, they are arguably easier to understand, and they  are `collapsible', i.e., the population marginal and conditional (on covariates or cluster effects or both) values are identical. Cluster-level analysis methods can be used to analyse CRTs where RD or RR is estimated as a measure of intervention effect \cite{hayes2009}, and these analyses can also incorporate adjustment for baseline covariates. These methods have the advantage of being simple to apply compared to the individual-level analysis methods. To date the performance of cluster-level analysis approaches with incompletely-observed binary outcomes has not been investigated.

The aim of this paper is two-fold. The first is to investigate the validity of estimating RD and RR as measures of intervention effect using unadjusted  and adjusted cluster-level analysis methods when binary outcomes are missing under a CDM mechanism. The second is to investigate the validity of individual-level analysis approaches considering the limitations of previous studies \cite{Ma2011, Ma2012comparing, Ma2013, Caille2014}, which we describe in Section \ref{Methods_handling_missing_data}. CRA and MI are used to handle the missing outcomes.

This paper is organised as follows. We begin in Section \ref{analysis_of_CRTs_with_complete_data} by giving a brief review of the approaches to the analysis of binary outcome in CRTs with full data. Section \ref{Methods_handling_missing_data} describes methods of handling missing data in CRTs. In Section \ref{validity_CRA},  we investigate the validity of CRA of CRTs under CDM assumption for missing binary outcomes. In Section \ref{sim_study}, we report the results of a simulation study to investigate the performance of our considered methods.  Section \ref{example} presents an example of application of our results to an actual CRT. We conclude in Section \ref{dis_con} with some discussion. 
\section{Analysis of CRTs with full data}
\label{analysis_of_CRTs_with_complete_data}
We begin by describing the two broad approaches to the analysis of CRTs in the absence of missing data. These two approaches are cluster-level analysis and individual-level analysis.  Let $ Y_{ijl} $ be a binary outcome of interest for the $ l $th $ (l=1,2,\ldots,m_{ij}) $ individual in the $ j $th $ (j=1,2,\ldots,k_i) $ cluster of the $ i $th $ (i=0,1) $ intervention group, where $ i=0 $ corresponds to control group and $ i=1 $ corresponds to intervention group. For convenience, we assume that both control and intervention groups have the same number of clusters $ (k_{i}=k) $ and constant cluster size across the groups $ (m_{ij}=m) $.  Also let $ X_{ijl} $ be an individual-level baseline covariate value for $ l $th individual in the $ (ij) $th cluster. Note that these methods can be extended to the case of multiple baseline covariates, some of which are individual-level and some are cluster-level.
\subsection{Cluster-level analysis}
This approach is conceptually very simple and can be explained as a two-stage process. Two different ways of doing  cluster-level analysis are unadjusted cluster-level analysis and (baseline covariate) adjusted cluster-level analysis. For binary outcomes, RD or RR is usually estimated as a measure of intervention effect  in cluster-level analysis \cite{hayes2009}. 
\subsubsection{Unadjusted cluster-level analysis ($ \text{CL}_\text{U} $)}:
\label{unadjusted_full}
In the first stage of analysis, a relevant summary measure of outcomes is obtained for each cluster. For binary outcomes, the cluster-level proportion of success is usually used as the summary measure for each cluster. Let $ p_{ij} $ be the observed proportion of successes in the $ (ij) $th cluster. Then RD is estimated as 
\begin{equation}\nonumber
\widehat{\text{RD}}_{\text{unadj}}=\bar{p}_1 -\bar{p}_0 
\end{equation} 
where $ \bar{p}_i $ is the mean of the cluster-specific proportions of success in the $ i $th intervention group. In the second stage, a test of the hypothesis $ \text{RD}=0 $ is performed using an appropriate statistical method.  The most popular one is the standard $ t- $test for two independent samples with degrees of freedom (DF)  $ 2k-2 $. The reason for using this test is that the cluster-specific summary measures are statistically independent, which is a  consequence of the clusters being independent of each other. 

Based on the first stage cluster level summary measures, RR is estimated as
\begin{equation}\nonumber
\widehat{\text{RR}}_{\text{unadj}}=\frac{\bar{p}_1}{\bar{p}_0}
\end{equation}  
Then, in the second stage, a test of the hypothesis $ \log\text{(RR)}=0 $ is performed using  $ t- $test with DF $ 2k-2 $, where  $ \widehat{\text{V}}\left( \log(\widehat{\text{RR}}_\text{unadj})\right) $ can be calculated as \cite{hayes2009} 
\begin{equation}\nonumber
\widehat{\text{V}}\left( \log(\widehat{\text{RR}}_\text{unadj})\right) \approx \frac{s_0^2}{k\bar{p}_0^2}+\frac{s_1^2}{k\bar{p}_1^2} \qquad \text{with}\qquad s_i^2=\frac{\sum_{j=1}^{k}\left(p_{ij}-\bar{p}_i \right)^2 }{k-1}
\end{equation} 
It can be shown that, with full data, $ \widehat{\text{RD}}_{\text{unadj}} $ is unbiased for  RD and $\widehat{\text{RR}}_{\text{unadj}}$ is consistent for RR as $ k \rightarrow \infty $ (Appendix \ref{app1}).
\subsubsection{Adjusted cluster-level analysis ($ \text{CL}_\text{A} $)}:
\label{adjusted_full}
In CRTs, baseline covariates that may be related to the outcome of interest are often collected and incorporated into the analysis. The main purpose of adjusting for covariates is to increase the credibility of the trial findings by demonstrating that any observed intervention effect is not attributable to the possible imbalance between the intervention groups in terms of baseline covariates \cite{Hernandez2004}.

In an adjusted cluster-level analysis, an individual-level regression analysis of the outcome of interest is carried out at the first stage of analysis ignoring the clustering of the data, which incorporates all covariates into the regression model except intervention indicator \cite{hayes2009,Gail1988}.   A standard logistic regression model is usually fitted for binary outcomes, which assumes that
\begin{equation}
\text{logit}\left( \pi_{ijl}\right)= \lambda_1+\lambda_2X_{ijl}
\label{model_firststage}
\end{equation}
where $ \pi_{ijl} $ is the probability that $ Y_{ijl} $ is 1. Let $ n_{ij} $ and $ \hat{n}_{ij} $ be the observed and predicted number of successes in the $ (ij) $th cluster, respectively. After fitting model (\ref{model_firststage}), $ \hat{n}_{ij} $ is calculated as
\begin{equation}\nonumber
\hat{n}_{ij} = \sum_{l=1}^{m}\hat{\pi}_{ijl} = \sum_{l=1}^{m} \text{expit} \left( \hat{\lambda}_1+\hat{\lambda}_2X_{ijl} \right) 
\end{equation}
where $ \text{expit}(t)= \exp(t)/(1+\exp(t)) $. Then the observed and predicted number of successes for each cluster are compared by computing a residual for each cluster.  In the case of no intervention effect, the residuals should be similar on average in the two intervention groups. 

If we want to estimate the adjusted RD, the residual, known as difference-residual,  for each cluster is calculated as 
\begin{equation}\nonumber
\epsilon^{d}_{ij}=\frac{n_{ij}-\hat{n}_{ij}}{m}
\end{equation}
where the  $ d $ superscript refers to difference-residual. The adjusted RD is then estimated as
\begin{eqnarray}
\widehat{\mbox{RD}}_{\text{adj}}&=&\bar{\epsilon}^{\,d}_1-\bar{\epsilon}^{\,d}_0\nonumber
\end{eqnarray}
where $ \bar{\epsilon}^{d}_i $ is the mean of the difference-residuals across the clusters  of intervention group $ i $, and where $ \widehat{\mbox{RD}}_{\text{adj}} $ can be rewritten as  
\begin{eqnarray}
\widehat{\text{RD}}_{\text{adj}}&=& \widehat{\text{RD}}_{\text{unadj}} + \frac{1}{mk}\sum_{j=1}^{k}\left( \hat{n}_{0j} - \hat{n}_{1j}  \right)\label{full_adj}
\end{eqnarray}
Since the distribution of $ X $ (in expectation) is the same between the intervention groups as a consequence of randomisation, and the prediction from the first-stage regression model (\ref{model_firststage}) depends only on $ X_{ijl} $,   $ \text{E}\left( \hat{n}_{0j}\right) = \text{E}\left( \hat{n}_{1j}\right) $. Hence, from (\ref{full_adj}),  $\widehat{\text{RD}}_{\text{adj}} $ is unbiased for RD since $ \widehat{\text{RD}}_{\text{unadj}} $ is unbiased for RD. In the second stage, a test of hypothesis $ \text{RD}_{\text{adj}}=0 $ is performed using $ t- $test with DF $ 2k-2 $. 

If we want to estimate the adjusted RR, the residual, also known as ratio-residual, for each cluster is calculated as 
\begin{equation}
\epsilon^{r}_{ij}=\frac{n_{ij}}{\hat{n}_{ij}}
\end{equation}
where the $ r $ superscript refers to ratio-residual. The adjusted RR is then estimated as
\begin{equation}
\widehat{\text{RR}}_{\text{adj}}=\frac{\bar{\epsilon}^{\,r}_1}{\bar{\epsilon}^{\,r}_0}\nonumber
\end{equation}
where $ \bar{\epsilon}^{\,r}_i $ is the mean of the ratio-residuals  across the clusters of intervention group $ i $. Then

\begin{eqnarray}
\text{E}\left( \widehat{\text{RR}}_{\text{adj}}\right) = \text{E}\left(  \frac{{\bar{\epsilon}^{\,r}_1}}{{\bar{\epsilon}^{\,r}_0}}\right)  \longrightarrow  \frac{\text{E} \left( \bar{\epsilon}^{\,r}_1 \right)}{\text{E} \left( \bar{\epsilon}^{\,r}_0 \right)} \text{~as~} k\rightarrow \infty 
\label{RRadj_consistent}
\end{eqnarray}
and
\begin{eqnarray}
\text{E}\left( \bar{\epsilon}^{\;r}_i\right) =\text{E}\left( \frac{1}{k}\sum_{j=1}^{k}\frac{n_{ij}}{\hat{n}_{ij}} \right)= \text{E}\left( \frac{n_{ij}}{\hat{n}_{ij}} \right) \longrightarrow  \frac{\text{E}\left( n_{ij}\right)}{\text{E}\left( \hat{n}_{ij}\right) } \text{~as~} m\rightarrow \infty  \nonumber
\end{eqnarray}
If $ \pi_i $ is the true proportion of success in the $ i $th intervention group, then 
\begin{equation}\nonumber
\text{E}\left( \bar{\epsilon}^{\;r}_i\right)\longrightarrow  \frac{\pi_i}{\text{E}\left( \hat{n}_{ij}/m\right) } \text{~as~} m\rightarrow \infty
\end{equation}
Hence, from equation (\ref{RRadj_consistent}), we can write
\begin{equation}\nonumber
\text{E}\left( \widehat{\text{RR}}_{\text{adj}}\right) \longrightarrow  \text{RR}\; \frac{\text{E}(\hat{n}_{0j})}{\text{E}(\hat{n}_{1j})} \text{~as~} (k, m)\rightarrow \infty
\end{equation}
As noted before, by randomisation $ \text{E}\left( \hat{n}_{0j}\right) = \text{E}\left( \hat{n}_{1j}\right) $.  Hence $ \widehat{\text{RR}}_{\text{adj}} $ is unbiased for true RR as $ (k,m)\to \infty $,  and therefore consistent. In the second stage, a test of hypothesis $ \log\left( \text{RR}_{\text{adj}}\right) =0 $ is performed using $ t- $test with DF $ 2k-2 $,  where  $ \widehat{\text{V}}\left( \log(\widehat{\text{RR}}_{\text{adj}})\right) $ can be calculated as 
\begin{equation}
\widehat{\text{V}}\left( \log(\widehat{\text{RR}}_\text{adj})\right) \approx \frac{s_{\epsilon 0}^2}{k\left( \bar{\epsilon}_0^r\right) ^2}+\frac{s_{\epsilon 1}^2}{k\left( \bar{\epsilon}_1^r\right) ^2} \qquad \text{with}\qquad s_{\epsilon i}^2=\frac{\sum_{j=1}^{k}\left(\epsilon_{ij}^r-\bar{\epsilon}_i^r \right)^2 }{k-1}
\label{var_adj_RR}
\end{equation} 
\subsection{Individual-level analysis} 
In individual-level analysis, a regression model is fitted to the individual-level outcome which allows us to analyse the effects of intervention and other covariates in the same model. For binary outcomes, two commonly used individual-level analysis methods are random-effects logistic regression (RELR), which estimates cluster-specific (also known as conditional) intervention effects, and generalised estimation equations (GEE), which estimates population-averaged (also known as marginal) intervention effects.  Both of these approaches are  extensions of the standard logistic regression models modified to allow for correlation between the outcomes of individuals in the same cluster.
\subsubsection{Random-effects logistic regression:}
Random-effects logistic regression (RELR) models take into account of between-cluster variability by incorporating cluster-specific random effects, which are almost always assumed to be normally distributed, into the logistic regression. These models are  fitted by maximising the likelihood function numerically, because the likelihood function and its derivative can not be derived analytically as this involves an integral over the distribution of the random effects.  Numerical integration methods are used to approximate the integral and so approximate the likelihood function. It is recommended to have at least 15 cluster in each intervention group to get the correct size and coverage for significance tests and confidence interval \cite{hayes2009}. Li and Redden \cite{Li2015} examined the performance of five denominator degrees of freedom (DDF) approximations, namely, residual DDF, containment DDF, between-within DDF, Satterthwaite DDF and Kenward-Roger DDF. They recommended to use between-within DDF approximation, which is equal to the total number of clusters in the study minus the rank of the design matrix, as it gave Type I error rate close to nominal level and higher power compared to the other four methods.  Ukoumunne \textit{et al.} \cite{Ukoumunne2007}  examined the properties of $ t- $based confidence intervals for log(OR) from CRTs  using degrees of freedom $ 2k-2 $ assuming the same number of clusters in the two intervention groups. They found that the coverage rates were close to the nominal level, although this approach gave overcoverage  with very small ICC (0.001). In this paper, we used the quantiles from $ t- $distribution with degrees of freedom $ 2k-2 $ rather than quantiles from $ \mathcal{N}(0,1) $ to construct the confidence interval for intervention effect. 
\subsubsection{Generalised estimating equations:}
Generalised estimating equations (GEE) are commonly used as a method for analysing binary outcomes in CRTs, while taking into account the correlation among the outcomes of the same cluster using a working correlation matrix.   In CRTs, it is usual to assume that the correlation matrix is exchangeable, since outcomes on individuals in different clusters are uncorrelated, while outcomes on individuals in the same cluster are equally correlated.  

In GEE, the sandwich standard error estimator is typically used to estimate the standard error of the parameter estimates. Although the sandwich standard error estimator is consistent even when the working correlation structure is specified incorrectly, the sandwich standard error of the regression coefficient tends to be biased downwards when the number of clusters in each intervention group is small \cite{hayes2009, Ukoumunne2007}. Moreover, the estimate of standard error is highly variable when the number of clusters is small. It is recommended to have at least 40 clusters in the study to get reliable standard error estimates \cite{Murray2004}. A number of methods have been proposed for dealing with the limitations of the sandwich variance estimator \cite{Mancl2001, Ukoumunne2007}. In this paper, we used the method proposed by Ukoumunne (2007) \cite{Ukoumunne2007} to correct the bias for small number of clusters in each intervention group. Firstly, the downward bias of the sandwich standard error estimator was  adjusted by multiplying it by $ \sqrt{k/(k-1)} $, where $ k $ is the number of clusters in each intervention group. Secondly, the increased small sample variability of the sandwich standard error estimator was accounted for by constructing the confidence interval for intervention effect based on the quantiles from a $ t- $distribution with degrees of freedom $ 2k-2 $ rather than quantiles from $ \mathcal{N}\left( 0,1 \right)  $. However, if some baseline covariates were cluster-level, the DF would be adjusted downwards as $ 2k-2-d $ to account for this, where $ d $ is the number of parameters corresponding to the cluster-level baseline covariates.   
\section{Methods of handling missing data in CRTs }
\label{Methods_handling_missing_data}
Common methods for handling missing data in CRTs are complete records analysis (CRA), single imputation and multiple imputation (MI).  In this paper, we focused on CRA and MI since they are the most commonly used methods for handling missing data. This section briefly describes these two approaches.
\subsection{Complete records analysis}
\label{CRA}
In complete records analysis (CRA), often referred to as complete case analysis,  only individuals with complete data on all variables in the analysis are considered. It has the advantage of being simple to apply, and is usually the default method in most statistical packages. It is well known that CRA is valid if data are MCAR. CRA is also valid if, conditioning on covariates, missingness is independent of outcome and and the outcome model being fitted is correctly specified \cite{littlerubin2002}. Based on simulations for CDM in CRTs, Ma \textit{et al.} \cite{Ma2012comparing, Ma2013} showed  that GEE using CRA performs well in terms of bias when the percentage of missing outcomes is low. In contrast, they concluded that  RELR using CRA dose not perform well. This is because they generated the data in such a way that they knew what the true population-averaged log(OR) was, but after fitting RELR, they compared estimates of conditional (on cluster random effects and covariates) log(OR) to the true population averaged log(OR). In addition, in the data generating mechanism used in these studies \cite{Ma2012comparing, Ma2013}, the baseline covariate was generated independently of the outcome, which in general is not a plausible assumption. It is therefore difficult to draw conclusions about what would happen in CRTs where the baseline covariates are related to the outcome. Caille \textit{et al.} \cite{Caille2014} reported  through simulations that GEE using unadjusted CRA and using adjusted (for covariates) CRA are biased for estimating intervention effects. However, in their simulation study, individual-level continuous outcomes were generated  at first using a linear mixed model which includes intervention indicator and a cluster random effect for each cluster, but without covariates. Each continuous outcome was  then dichotomised to obtain a binary outcome. Then, baseline covariates were generated dependent on the continuous outcomes. So it appears the data generation mechanism used would mean that baseline covariates were associated with intervention group, which is not possible (in expectation) due to randomisation. In addition, as the authors noted, they compared estimates of covariate conditional ORs to the true unconditional ORs, which would be expected to differ even with full data due to non-collapsibility. It is therefore difficult to draw general conclusions from their results about the methods' performance in CRTs.  
\subsection{Multiple Imputation}
\label{MI}
In multiple imputation (MI) method, a sequence of $ N $ imputed data sets are obtained by replacing each missing outcome by a set of $ N\ge 2 $ imputed values that are simulated from an appropriate distribution or model. Imputing multiple times allows the uncertainty associated with the imputed values due to the fact that the imputed values are sampled draws for the missing outcomes instead of the actual values. This uncertainty is taken into account by adding between-imputation variance to the average within-imputation variance. Each of the $ N $ imputed data sets are analysed as a full data set using standard methods and the results are then combined using Rubin's rules \cite{rubin1987}. One important feature of MI is that the imputation model and the analysis model do not have to be the same. However, in order for Rubin's rules to be valid, the imputation model needs to be compatible or congenial with the analysis model in the sense that the imputation model has to ``contain" the analysis model \cite{meng1994}.   

There are at least four different types of MI that have been used in CRTs \cite{DiazOrdaz2014}. These are \textit{standard} MI, also known as \textit{single-level} MI,  which ignores clustering in the imputation model, \textit{fixed effects} MI  which includes a fixed effect for each cluster in the imputation model, \textit{random effects} MI where clustering is taken into account through a random effect for each cluster in the imputation model and \textit{within-cluster} MI where standard MI is applied within each cluster. From now, we refer to random effects MI as multilevel multiple imputation (MMI).   

The multiple imputation inference is usually based on a $ t- $ distribution with degrees of freedom (DF) given by
\begin{equation}
\upsilon=(N-1)\left( 1+\frac{N}{N+1} \frac{W}{B}\right)^2
\label{df_com}
\end{equation} 
where $ B $ and $ W $  are the between-imputation variance and the average  within-imputation variance, respectively. This DF is derived under the assumption that the complete data (full data) DF, $ \upsilon_\text{com} $, is infinite \cite{Barnard1999}. In CRTs,  the value of $ \upsilon_{\text{com}} $ is calculated based on the number of clusters in the study rather than the number of individuals and, therefore, is usually small. In CRTs with equal number of clusters in each intervention group, $ \upsilon_{\text{com}} $ is calculated as $ 2k-2 $ \cite{Andridge2011}. If $ \upsilon_{\text{com}} $ is small and there is a modest proportion of missing data,  the value of $ \upsilon $  can be much higher than $ \upsilon_{\text{com}} $, which is not appropriate \cite{Barnard1999}. In such a situation, a more appropriate DF, proposed by Barnard and Rubin (1999) \cite{Barnard1999}, is calculated as
\begin{equation}
\nu_{\text{adj}}=\left({\upsilon}^{-1} +\hat{\upsilon}_{\text{obs}}^{-1}\right)^{-1} \le  \nu_{\text{com}}
\label{df.adj} 
\end{equation}
where
\begin{equation} \nonumber
\hat{\nu}_{\text{obs}}=  \left( \frac{\nu_{\text{com}}+1}{\nu_{\text{com}}+3}\right)\nu_{\text{com}} \left( 1+\frac{N+1}{N} \frac{B}{W} \right)^{-1} 
\end{equation}

Ma \textit{et al.} \cite{Ma2011} examined  within-cluster MI, fixed effects MI and MMI for missing binary outcomes under CDM mechanism in CRTs.  They showed that all these strategies give quite similar results for low percentages of missing data or for small value of ICC. With high percentage of missing data, the within-cluster MI  underestimate the variance of the intervention effect which may result in inflated Type I error rate. In two subsequent studies, Ma \textit{et al.} \cite{Ma2012comparing, Ma2013} compared the performance of GEE and RELR with missing binary outcomes using standard MI and within-cluster MI. Results showed that GEE  performs well when using standard MI and the variance inflation factor (VIF) is less than 3; and using within-cluster MI  when VIF $ \ge 3$ and cluster size is at least 50. Ma \textit{et al.} \cite{Ma2013} concluded that RELR does not perform well using either standard MI or within-cluster MI. However, in the later two studies \cite{Ma2012comparing, Ma2013}, as we described in Section \ref{CRA}, they compared estimates of conditional (on cluster random effects and covariates) log(OR) to the true population averaged log(OR); and their data generation mechanisms do not corresponds to how data arise in CRTs. In the first study \cite{Ma2011}, the simulation was based on a real dataset, so the conclusions to other design settings may be limited. It is therefore again difficult to draw conclusions from their results about the performance of GEE and RELR with different MI strategies under CDM mechanism. Caille \textit{et al.} \cite{Caille2014} compared different MI strategies through a simulation study for handing missing binary outcomes in CRTs assuming CDM, assessing bias, standard error and coverage rate of the estimated intervention effect. They showed that MMI with RELR and single-level MI with standard logistic regression give better inference for intervention effect compared to CRA in terms of bias, efficiency and coverage. However, as we described in Section \ref{CRA}, their data generation mechanism does not correspond to how data arise in CRTs. It is therefore again difficult to draw general conclusions from their results about the MI strategies' performance in CRTs.

In the case of missing continuous outcome in CRTs, Andridge \cite{Andridge2011} showed that the true MI variance of group means are underestimated by single-level MI, and are overestimated by fixed effects MI. She also showed that MMI is the best among these three methods and recommended its use for practitioners. Diaz-Ordaz \textit{et al.} \cite{Diazordaz2016} showed that for bivariate outcomes MMI gives coverage rate close to nominal level,  whereas single-level MI gives low coverage and fixed effects MI gives overcoverage. In this paper, we therefore used MMI for missing binary outcome.

\section{Validity of CRA of CRTs}
\label{validity_CRA}
In this section, we investigate the validity of $ \text{CL}_\text{U} $, $ \text{CL}_\text{A} $, RELR and GEE using complete records, when binary outcomes are missing under CDM. 

In settings where the expectation of the outcome is assumed to be linearly dependent on the covariate and intervention indicator, both unadjusted  and adjusted cluster-level analyses for estimating mean difference as a measure of intervention effect are unbiased in general only when the two intervention groups have the same CDM mechanism and the same covariate effect on the outcome \cite{Hossain2016}. However, the assumption of the expectation of the outcome being linear in baseline covariate  and intervention indicator is not very plausible in the case of a binary outcome. Two common alternatives to assuming the expectation is a linear function of baseline covariates and intervention indicator are to use a log or logit link between the mean of the outcome and the linear predictor.    

Assuming the true data generating model has log link, suppose that each binary outcome $ Y_{ijl} $ is generated by 
\begin{equation}
\pi_{ijl}=\exp(\beta_0+\beta_1 i+f_{i}(X_{ijl})+\delta_{ij})
\label{model_loglink}
\end{equation}
where   $ \beta_0 $ is a constant, $ \beta_1 $ is the true intervention effect, $ f_{i}(X_{ijl}) $ is a function of baseline covariate $ X $ in the $ i $th intervention group, $ \delta_{ij} $ is the $ (ij) $th cluster effect with mean 0 and variance $ \sigma_{b}^2 $, and $ \pi_{ijl}=P\left( Y_{ijl}=1| \delta_{ij},X_{ijl}\right)  $. If on the other hand, we assume a logit link for the true data generating model, we have
\begin{equation}
\pi_{ijl}=\text{expit}\left( \beta_0+\beta_1 i+f_{i}(X_{ijl})+\delta_{ij}\right) 
\label{model_logitlink}
\end{equation}
Define a missing data indicator $ R_{ijl} $ such that
\begin{equation}
R_{ijl}=\begin{cases}
1, & \text{if~} Y_{ijl} \text{~ is observed}\\
0, & \text{if~} Y_{ijl} \text{~ is missing}
\end{cases}
\label{miss_indicator}
\end{equation}
Then $ \sum_{l=1}^{m}R_{ijl} $ is the number of complete records in the $ (ij) $th cluster. \subsection{Cluster-level analyses for estimating RD}
\label{validity_RD}
In unadjusted cluster-level analysis using complete records, RD is estimated as
\begin{equation}
\widehat{\text{RD}}_{\text{unadj}}^{\text{cr}} =\bar{p}^{\text{~cr}}_{1}-\bar{p}^{\text{~cr}}_{0}\label{rd_cra} 
\end{equation}
where $ \bar{p}^{\text{~{cr}}}_{i} $ is the mean of the cluster-specific proportions of success, calculated using complete records, in the $ i $th intervention group. The superscript \textbf{cr} refers to complete records.

In adjusted cluster-level analysis, recall that a logistic regression model is fitted to the data at the first stage of analysis ignoring intervention and clustering of the data. Then the observed and predicted number of successes in each cluster are compared by computing a residual for each cluster. The adjusted RD using complete records is estimated as 
\begin{eqnarray}
\widehat{\text{RD}}_{\text{adj}}^{\text{cr}} &=&\bar{\epsilon}_1^{\,d\text{(cr)}}-\bar{\epsilon}_0^{\,d\text{(cr)}}
\label{estimate_RD_CCA}
\end{eqnarray}
where $ \bar{\epsilon}_i^{\,d\text{(cr)}}$ is the average of the cluster-specific difference-residuals in the $ i $th intervention group using complete records, which are calculated as 
\begin{equation}\nonumber
\epsilon_{ij}^{d\text{(cr)}}=\frac{n_{ij}^{\text{cr}}-\hat{n}_{ij}^{\text{cr}}}{\sum_{l=1}^{m}R_{ijl}}
\end{equation}
where $ n_{ij}^{\text{cr}} $ and $\hat{n}_{ij}^{\text{cr}}  $ are the observed and predicted number of successes in the $ (ij) $th cluster using complete records. Then $ \widehat{\text{RD}}_{\text{adj}}^{\text{cr}}  $ can be written in terms of $ \widehat{\text{RD}}_{\text{unadj}}^{\text{cr}}  $ as
\begin{eqnarray}
\widehat{\text{RD}}_{\text{adj}}^{\text{cr}} 
&=& \widehat{\text{RD}}_{\text{unadj}}^{\text{cr}} + \frac{1}{k}\sum_{j=1}^{k} \left[ \frac{1}{\sum_{l=1}^{m}R_{ijl}}\left(\hat{n}_{0j}^{\text{cr}} - \hat{n}_{1j}^{\text{cr}} \right) \right] 
\label{RDadj_CRA}
\end{eqnarray}

We aim to derive conditions under which the cluster-level analyses for RD using complete records are unbiased. To this end, we write the individual-level probabilities of success, $ \pi_{ijl} $, as 
\begin{equation}
\pi_{ijl}=\pi_i+g_i\left( X_{ijl}, \delta_{ij}\right) 
\end{equation}
where $ g_i\left( X_{ijl}, \delta_{ij}\right) $ is a function of baseline covariate $ X_{ijl} $ and random cluster-effect $ \delta_{ij} $, and which determines how individual-level probabilities of success differ from group level probability of success in each intervention group. Then
\begin{equation}
\text{E}_{j,l}\left( \pi_{ijl}|R_{ijl}=1\right) = \pi_i + \text{E}_{j,l}\left( g_i\left( X_{ijl}, \delta_{ij}\right)| R_{ijl}=1 \right)\nonumber
\end{equation}
and 
\begin{eqnarray}
\text{E}\left( \widehat{\text{RD}}_{\text{unadj}}^{\text{cr}}\right) &=& \text{E}\left( \pi_{1jl}|R_{1jl}=1\right)-\text{E}\left( \pi_{0jl}|R_{0jl}=1\right)\nonumber\\
&=& \pi_1-\pi_0 + \text{E}\left( g_1\left( X_{1jl}, \delta_{1j}\right)| R_{1jl}=1 \right)-\text{E}\left( g_0\left( X_{0jl}, \delta_{0j}\right)|R_{0jl}=1 \right) \nonumber\\
&=&\text{RD} + \text{E}\left( g_1\left( X_{1jl}, \delta_{1j}\right)| R_{1jl}=1 \right)-\text{E}\left( g_0\left( X_{0jl}, \delta_{0j}\right)| R_{0jl}=1 \right) \nonumber
\end{eqnarray}
So $  \widehat{\text{RD}}_{\text{unadj}}^{\text{cr}} $ will be unbiased for true RD if and only if 
\begin{equation}
\text{E}\left( g_1\left( X_{1jl}, \delta_{1j}\right)| R_{1jl}=1 \right)= \text{E}\left( g_0\left( X_{0jl}, \delta_{0j}\right)|  R_{0jl}=1 \right)
\label{conditon1}
\end{equation}
Assuming  the data are generated from a log link model (\ref{model_loglink}), we have 
\begin{eqnarray}
g_i\left(X_{ijl},\delta_{ij} \right)&=&\pi_{ijl}-\pi_i \nonumber\\
&=& \exp(\beta_0+\beta_1 \,i)\left\lbrace \exp\left( f_{i}(X_{ijl})+\delta_{ij}\right)-\text{E}_{j,l}\left( \exp\left(f_{i}(X_{ijl})+\delta_{ij}\right)\right)   \right\rbrace
\label{log_RD_condition}
\end{eqnarray}
since $ \pi_i=\text{E}_{j,l}\left(\pi_{ijl} \right)  $. If there is an intervention effect in truth $ (\beta_1\ne 0) $, in general, we have from (\ref{log_RD_condition})
\begin{equation}
\text{E}\left( g_1\left( X_{1jl}, \delta_{1j}\right)| R_{1jl}=1\right) \ne \text{E}\left( g_0\left( X_{0jl}, \delta_{0j}\right)| R_{0jl}=1\right)\nonumber
\end{equation}
even if the two intervention groups have the same missingness mechanism and the same covariate effects in the data generating model for the outcome. Hence, $\widehat{\text{RD}}_{\text{unadj}}^{\text{cr}} $ is biased for true RD when the true data generating model has log link.  However, under the null hypothesis of no intervention effect $ (\beta_1=0) $, if the two intervention groups have the same covariate effect, i.e.  $ f_{i}(X_{ijl})=f(X_{ijl}) \text{ for } i \in \{0,1\} $, we have
\begin{equation}
g_i\left(X_{ijl},\delta_{ij} \right)
= \exp(\beta_0)\left\lbrace \exp\left( f(X_{ijl})+\delta_{ij}\right)-\text{E}_{j,l}\left( \exp\left(f(X_{ijl})+\delta_{ij}\right)\right)   \right\rbrace \nonumber
\end{equation}
and then, in addition, if the two intervention groups have the same missingness mechanism, we have
\begin{equation}
\text{E}\left( g_1\left( X_{1jl}, \delta_{1j}\right)| R_{1jl}=1\right) = \text{E}\left( g_0\left( X_{0jl}, \delta_{0j}\right)| R_{0jl}=1\right)\nonumber
\end{equation}
and hence $  \widehat{\text{RD}}_{\text{unadj}}^{\text{cr}} $ is unbiased for true $ \text{RD}=0 $.

On the other hand, if we assume the data are generated from a logit link model (\ref{model_logitlink}), we have 
\begin{eqnarray}
g_i\left(X_{ijl},\delta_{ij} \right)&=&\pi_{ijl}-\pi_i\nonumber \\
&=& \text{expit}\left( \beta_0+\beta_1 i+f_{i}(X_{ijl})+\delta_{ij}\right)  -\text{E}_{j,l}\left( \text{expit}\left( \beta_0+\beta_1 i+f_{i}(X_{ijl})+\delta_{ij}\right)\right)
\label{logit_RD_condition}
\end{eqnarray} 
Then, again with $ \beta_1\ne 0 $,
\begin{equation}
\text{E}\left( g_1\left( X_{1jl}, \delta_{1j}\right)| R_{1jl}=1\right) \ne \text{E}\left( g_0\left( X_{0jl}, \delta_{0j}\right)| R_{0jl}=1\right)\nonumber
\end{equation}  
even if the two intervention groups have the same missingness mechanism and the same covariate effect. Hence, $  \widehat{\text{RD}}_{\text{unadj}}^{\text{cr}} $ is biased for true RD when the true data generating model has logit link. However, like log link, under the null hypothesis of no intervention effect $ (\beta_1=0) $, if the two intervention groups have the same covariate effect, i.e.  $ f_{i}(X_{ijl})=f(X_{ijl}) \text{ for } i \in \{0,1\} $, we have
\begin{equation}
g_i\left(X_{ijl},\delta_{ij} \right)= \text{expit}\left( \beta_0+f(X_{ijl})+\delta_{ij}\right)  -\text{E}_{j,l}\left( \text{expit}\left( \beta_0+f(X_{ijl})+\delta_{ij}\right) \right) \nonumber
\end{equation} 
and then, in addition, if the two intervention groups have the same missingness mechanism, we have
\begin{equation}
\text{E}\left( g_1\left( X_{1jl}, \delta_{1j}\right)| R_{1jl}=1\right) = \text{E}\left( g_0\left( X_{0jl}, \delta_{0j}\right)| R_{0jl}=1\right)\nonumber
\end{equation}
and hence $  \widehat{\text{RD}}_{\text{unadj}}^{\text{cr}} $ is unbiased for true $ \text{RD}=0 $.

Referring to equation (\ref{RDadj_CRA}), if the two intervention groups have the same missingness mechanism and the same covariate effect, then $\text{E}\left( \hat{n}_{0j}^{\text{cr}}\right) = \text{E}\left( \hat{n}_{1j}^{\text{cr}}\right) $. Hence, with $ \beta_1\ne 0 $, from equation (\ref{RDadj_CRA}), we can conclude that since $ \widehat{\text{RD}}_{\text{unadj}}^{\text{cr}} $ is biased with both log and logit links for the true data generating model,  $ \widehat{\text{RD}}_{\text{adj}}^{\text{cr}}  $ is also biased for RD with both log and logit links in the true data generating model. However, with $ \beta_1=0 $, since $ \widehat{\text{RD}}_{\text{unadj}}^{\text{cr}}  $ is unbiased for RD with both log and logit links, when the two intervention groups have the same missingness mechanism and the same covariate effect, $ \widehat{\text{RD}}_{\text{adj}}^{\text{cr}}  $ is also unbiased for RD under the same conditions. It can also be seen from (\ref{log_RD_condition}) and (\ref{logit_RD_condition}) that
\begin{equation}
\text{E}_{j,l}\left( g_i\left( X_{ijl}, \delta_{ij}\right)\right) = 0 \text{ for } i \in \{0,1\}\nonumber
\end{equation}
for both log and logit links in the data generating model, and hence both $ \widehat{\text{RD}}_{\text{adj}}  $ and $ \widehat{\text{RD}}_{\text{adj}}  $ are unbiased for true RD with full data. 
\subsection{Cluster-level analyses for estimating RR}
\label{validity_RR}
In unadjusted cluster-level analysis using complete records, RR is estimated as
\begin{equation}
\widehat{\text{RR}}_{\text{unadj}}^{\text{cr}}=\frac{\bar{p}^{\text{~cr}}_{1}}{\bar{p}^{\text{~cr}}_{0}}\nonumber
\end{equation}
and, in adjusted cluster-level analysis, the adjusted RR using complete records is estimated as
\begin{equation}
\widehat{\text{RR}}_{\text{adj}}^{\text{cr}} =\frac{\bar{\epsilon}_1^{\,r\text{(cr)}}}{\bar{\epsilon}_0^{\,r\text{(cr)}}}
\label{RRadj_CRA}
\end{equation}
where $\bar{\epsilon}_i^{\,r\text{(cr)}}$ is the average of the ratio-residuals in the $ i $th intervention group using complete records, which are calculated as 
\begin{equation}\nonumber
\epsilon_{ij}^{r\text{(cr)}}=\frac{n_{ij}^{\text{cr}}}{\hat{n}_{ij}^{\text{cr}}}\\
\end{equation}
We aim to establish conditions under which the cluster-level analyses for RR using complete records are consistent. To this end, we write  $ \pi_{ijl} $  as 
\begin{equation}
\pi_{ijl}=\pi_i \;h_i\left( X_{ijl}, \delta_{ij}\right) 
\end{equation}
where $ h_i\left( X_{ijl}, \delta_{ij}\right) $ is a function of baseline covariate $ X_{ijl} $ and random cluster-effect $ \delta_{ij} $, and which determines how individual-level probabilities of success differ from group level probability of success. Then
\begin{equation}
\text{E}_{j,l}\left( \pi_{ijl}|R_{ijl}=1\right) = \pi_i \; \text{E}_{j,l}\left( h_i\left( X_{ijl}, \delta_{ij}\right)|R_{ijl}=1 \right)\nonumber
\end{equation}
and 
\begin{eqnarray}
\text{E}\left( \widehat{\text{RR}}_{\text{unadj}}^{\text{cr}}\right) & \longrightarrow  & \frac{\text{E}\left( \pi_{1jl}| R_{1jl}=1\right)}{\text{E}\left( \pi_{0jl}| R_{0jl}=1\right)} \text{ as } (k,m) \longrightarrow \infty \nonumber\\
&=& \frac{\pi_1\; \text{E}\left( h_1\left( X_{1jl}, \delta_{1j}\right)| R_{1jl}=1 \right)}{\pi_0\;\text{E}\left( h_0\left( X_{0jl}, \delta_{0j}\right)| R_{0jl}=1 \right)} \nonumber\\
&=&\text{RR} \; \frac{\text{E}\left( h_1\left( X_{1jl}, \delta_{1j}\right)|  R_{1jl}=1 \right)}{\text{E}\left( h_0\left( X_{0jl}, \delta_{0j}\right)| R_{0jl}=1 \right)} \nonumber
\end{eqnarray}
So $  \widehat{\text{RR}}_{\text{unadj}}^{\text{cr}} $ will be consistent for true RR if only if
\begin{equation}
\frac{\text{E}\left( h_1\left( X_{1jl}, \delta_{1j}\right)| R_{1jl}=1 \right)}{\text{E}\left( h_0\left( X_{0jl}, \delta_{0j}\right)|  R_{0jl}=1 \right)}=1
\label{conditon2}
\end{equation}
Assuming  the data are generated from a log link model (\ref{model_loglink}), we have 
\begin{eqnarray}
h_i\left(X_{ijl},\delta_{ij} \right)&=&\frac{\exp\left( \beta_0+\beta_1 i + f_i(X_{ijl})+\delta_{ij}\right)}{\text{E}_{j,l}\left( \exp\left(\beta_0+\beta_1 i +f_i(X_{ijl})+\delta_{ij}\right)\right)} \nonumber\\
&=&  \frac{\exp\left( f_i(X_{ijl})+\delta_{ij}\right)}{\text{E}_{j,l}\left( \exp\left(f_i(X_{ijl})+\delta_{ij}\right)\right)} \nonumber
\label{log_RR_condition}
\end{eqnarray}
and
\begin{equation*}
\frac{\text{E}\left( h_1\left( X_{1jl},\delta_{1j}\right)| R_{1jl}=1\right)}{\text{E}\left( h_0\left( X_{0jl},\delta_{0j}\right)| R_{0jl}=1\right)}= \frac{\text{E}\left( \exp\left( f_1(X_{1jl})+\delta_{1j}\right)| R_{1jl}=1\right) }{\text{E}\left( \exp\left( f_0(X_{0jl})+\delta_{0j}\right)| R_{0jl}=1\right)}
\times\frac{\text{E}\left( \exp\left(f_0(X_{0jl})+\delta_{0j}\right) \right)}{\text{E}\left( \exp\left(f_1(X_{1jl})+\delta_{1j}\right) \right)} \nonumber
\end{equation*}
Then if the two intervention groups have the same covariate effect, i.e. $ f_i(X_{ijl})=f(X_{ijl}) \text{ for } i\in \{0,1\} )$, we have
\begin{equation}
\frac{\text{E}\left( \exp\left(f_0(X_{0jl})+\delta_{0j}\right)\right)}{\text{E}\left( \exp\left(f_1(X_{1jl})+\delta_{1j}\right)\right)}=1 \nonumber
\end{equation}
and, in addition, if the two intervention groups have the same missingness mechanism, we have
\begin{equation}
\frac{\text{E}\left( \exp\left( f_1(X_{1jl})+\delta_{1j}\right)| R_{1jl}=1\right) }{\text{E}\left( \exp\left( f_0(X_{0jl})+\delta_{0j}\right)| R_{0jl}=1\right)}=1 \nonumber
\end{equation} 
Therefore, if the two intervention groups have the same missingness mechanism and the same covariate effects, we have
\begin{equation}
\frac{\text{E}\left( h_1\left( X_{1jl},\delta_{1j}\right)| R_{1jl}=1\right)}{\text{E}\left( h_0\left( X_{0jl},\delta_{0j}\right)| R_{0jl}=1\right)}=1\nonumber
\end{equation}
and hence  $  \widehat{\text{RR}}_{\text{unadj}}^{\text{cr}} $ is consistent for true RR. 

On the other hand, assuming the data are generated from logit link model (\ref{model_logitlink}), we have 
\begin{eqnarray}
h_i\left(X_{ijl},\delta_{ij} \right)
&=& \frac{\text{expit}\left( \beta_0+\beta_1 i+f_i(X_{ijl})+\delta_{ij}\right)}{\text{E}_{j,l}\left( \text{expit}\left( \beta_0+\beta_1 i+f_i(X_{ijl})+\delta_{ij}\right)\right)} \nonumber
\end{eqnarray}
and
\begin{multline}
\frac{\text{E}\left( h_1\left( X_{1jl},\delta_{1j}\right)| R_{1jl}=1\right)}{\text{E}\left( h_0\left( X_{0jl},\delta_{0j}\right)| R_{0jl}=1\right)}= \frac{\text{E}\left( \text{expit}\left( \beta_0+\beta_1+f_1(X_{1jl})+\delta_{1j}\right)| R_{1jl}=1\right) }{\text{E}\left( \text{expit}\left(\beta_0+ f_0(X_{0jl})+\delta_{0j}\right)| R_{0jl}=1\right)}\\
\times\frac{\text{E}\left( \text{expit}\left(\beta_0+f_0(X_{0jl})+\delta_{0j}\right) \right)}{\text{E}\left( \text{expit}\left(\beta_0+\beta_1+f_1(X_{1jl})+\delta_{1j}\right) \right)}
\label{result_logit}
\end{multline}
If $ \beta_1\ne 0 $, we have
\begin{equation}
\frac{\text{E}\left( \text{expit}\left(\beta_0+f_0(X_{0jl})+\delta_{0j}\right) \right)}{\text{E}\left( \text{expit}\left(\beta_0+\beta_1+f_1(X_{1jl})+\delta_{1j}\right) \right)}\ne 1 \nonumber
\end{equation}
and
\begin{equation}
\frac{\text{E}\left( \text{expit}\left( \beta_0+\beta_1+f_1(X_{1jl})+\delta_{1j}\right)| R_{1jl}=1\right) }{\text{E}\left( \text{expit}\left(\beta_0+ f_0(X_{0jl})+\delta_{0j}\right)| R_{0jl}=1\right)} \ne 1 \nonumber
\end{equation} 
even if the two intervention groups have the same missingness mechanism and the same covariate effects. Hence
\begin{equation}
\frac{\text{E}\left( h_1\left( X_{1jl},\delta_{1j}\right)| R_{1jl}=1\right)}{\text{E}\left( h_0\left( X_{0jl},\delta_{0j}\right)| R_{0jl}=1\right)}\ne 1 \nonumber
\end{equation}
and therefore $  \widehat{\text{RR}}_{\text{unadj}}^{\text{cr}} $ is not consistent for true RR. However, under the null hypothesis of no intervention effect $ (\beta_1=0) $, if the two intervention group have the same missingness mechanism and the same covariate effect, the both ratios of expectations in the right side of equation (\ref{result_logit}) equal to one, and hence we have
\begin{equation}
\frac{\text{E}\left( h_1\left( X_{1jl},\delta_{1j}\right)| R_{1jl}=1\right)}{\text{E}\left( h_0\left( X_{0jl},\delta_{0j}\right)| R_{0jl}=1\right)} = 1 \nonumber
\end{equation}
Therefore, if the data generating model has logit link and there is no intervention effect in truth, $  \widehat{\text{RR}}_{\text{unadj}}^{\text{cr}} $ is consistent for true $ \text{RR}=1 $ when the two intervention groups have the same missingness and the same covariate effect. 

From equation (\ref{RRadj_CRA}), we can write
\begin{eqnarray}
\text{E}\left( \widehat{\text{RR}}_{\text{adj}}^{\text{cr}}\right) 
&\longrightarrow & \text{E} \left( \widehat{\text{RR}}_{\text{unadj}}^{\text{cr}} \right) \frac{\text{E}\left(\hat{p}_{0j}^{\text{cr}} \right) }{\text{E}\left(\hat{p}_{1j}^{\text{cr}} \right)} \text{ as }(k,m)\to \infty. \quad
(\text{ see Appendix \ref{Appendix B}})
\label{equation_RRadj}
\end{eqnarray}
If the two intervention groups have the same missingness mechanism and the same covariate effect, then  $\text{E}\left(\hat{p}_{0j}^{\text{cr}} \right)=\text{E}\left(\hat{p}_{1j}^{\text{cr}} \right)$ since the distribution of $ X $ is the same (by randomisation) across the intervention groups. As we have already shown that $ \widehat{\text{RR}}_{\text{unadj}}^{\text{cr}}  $ is consistent for RR with log link in the true data generating model, when the two intervention groups have the same missingness mechanism and the same covariate effects, $ \widehat{\text{RR}}_{\text{adj}}^{\text{cr}} $ is also consistent for RR under the same conditions. Similarly, in the presence of a true intervention effect, since $\widehat{\text{RR}}_{\text{unadj}}^{\text{cr}} $ is not consistent for RR if the true data generating model has a logit link, $ \widehat{\text{RR}}_{\text{adj}}^{\text{cr}} $ is also not consistent for RR under the same conditions.  However, under the null hypothesis, since $\widehat{\text{RR}}_{\text{unadj}}^{\text{cr}} $ is consistent for $ \text{RR}=1 $ if the true data generating model has logit link and the two intervention groups have the same missingness mechanism and the same covariate effects, $ \widehat{\text{RR}}_{\text{adj}}^{\text{cr}} $ is also consistent for $ \text{RR}=1 $ under the same conditions.
\subsection{RELR and GEE using complete records}
For individually randomised trials, it is well known that likelihood based CRA is valid under MAR, if missingness is only in the outcome and all predictors of missingness are included in the model as covariates \cite{littlerubin2002}.  So  it is anticipated that RELR using CRA will give consistent estimates of intervention effect, if the covariate $ X $, which is associated with the missingness, is included in the model and the model is correctly specified.  We also expect that GEE  using CRA adjusted for covariate $ X $ which is associated with the missingness in outcomes will give consistent estimates of intervention effect. 

When it is assumed that the two intervention groups have the same covariate effects on outcome, we fit RELR with fixed effects of intervention indicator and covariate, and a random effect for cluster; while we fit GEE with intervention indicator and covariate assuming exchangeable correlation for the outcomes of the same cluster.  If it is assumed that the baseline covariate effect on outcome could be different in the two intervention groups, an interaction between intervention and covariate must be included in the model. This implies that the intervention effect varies with level of covariate values. In those scenarios where an interaction is present, we will target the intervention effect at the mean value of the covariate. Let $ X^* $ denote the empirically centred covariate $ X-\bar{X} $, where $ \bar{X} $ is the mean of $ X $ using data from all individuals. Then,  we fit RELR with fixed effects of intervention indicator,  $ X^* $ and their interaction, and a random effect for cluster; while we fit GEE including the intervention indicator, $X^* $ and their interaction, and assuming an exchangeable correlation for the outcomes of the same cluster. One may need to account for the centreing step in the variance estimation. We will investigate in the simulation whether ignoring this has any negative impact on confidence interval coverage. 

\section{Simulation Study}
\label{sim_study}
A simulation study was conducted to assess the performance of $ \text{CL}_\text{U} $, $ \text{CL}_\text{A} $,  RELR and GEE under CDM mechanism. CRA and multilevel multiple imputation (MMI) were used to handle the missing data. The average estimate of intervention effect, its average estimated standard error (SE) and coverage rates were calculated for each of the methods and compared to each other.  We considered balanced CRTs, where the two intervention groups have the same number of clusters and constant cluster size (before missing outcomes were introduced), and a single continuous individual-level baseline covariate. 
\subsection{Data generation}
Data were generated using the model in equation (\ref{model_logitlink}) with a logit link, as described in Section \ref{validity_CRA}, with $ f_i(X_{ijl})=\beta_{2(i)}X_{ijl} $, where $ \beta_{2(i)} $ is the effect of covariate of $ X $ in the $ i $th intervention group. Firstly, for each individual in the study, a value of  $ X_{ijl} $ was generated using the model
\begin{equation} \nonumber
X_{ijl}=\alpha_{ij} +u_{ijl}
\end{equation} 
where $ \alpha_{ij} $ is the $ (ij) $th cluster effect on $ X $ and  $ u_{ijl} $ is the individual-level error on $ X $. We assumed that $  \alpha_{ij}\sim \mathcal{N}\left( \mu_x,\sigma_{\alpha}^2  \right)$  independently of   $ u_{ijl}\sim \mathcal{N}\left(  0,\sigma_u^2 \right) $, where $ \mu_{x} $ is the mean of $ X $, $ \sigma_{\alpha}^2 $ and $ \sigma_{u}^2 $ are the between-cluster and within-cluster variance of $ X $, respectively.  The total variance of $ X $ can be written as  $ \sigma_x^2 =\sigma_{\alpha}^2+\sigma_u^2 $ and thus the ICC of $ X $ is $ \rho_{x}=\sigma_{\alpha}^2/\sigma_{x}^2 $. Then we generated $ \text{logit}(\pi_{ijl}) $ for each individual in the study using model (\ref{model_logitlink}) assuming $ \delta_{ij}\sim \mathcal{N}\left(  0,\sigma_b^2 \right) $. Finally, $ Y_{ijl} $ was generated as Bernoulli random variable with parameter $ \pi_{ijl} $.  Without loss of generality, we set $ \beta_0=0 $ in equation (\ref{model_logitlink}) for convenience and chose the others parameters   to obtain pre-specified value of success rates $ \pi_0 $ and $ \pi_1 $ in the control and intervention groups, respectively, on average over 1000 data sets. We varied the number of clusters in each intervention group as $ k=(5,10,20,50) $ and fixed the cluster size $ m=50 $. 

Once the complete data (full data) sets were generated, we introduced missing outcomes by generating a  missing outcome data indicator $ R_{ijl} $ (defined in equation (\ref{miss_indicator})), independently for each individual, under covariate dependent missingness (CDM) mechanism according to a logistic regression model
\begin{equation}
\text{logit}(R_{ijl}=0|\boldsymbol{Y}_{ij},\boldsymbol{X_{ij}})=\psi_{i}+\phi_{i}X_{ijl}
\label{miss}
\end{equation}
where $ \boldsymbol{Y}_{ij} $ and $ \boldsymbol{X}_{ij} $ are the vectors of outcome and covariate values, respectively, of the $ (ij) $th cluster. The constants $ \psi_{i} $ and $ \phi_{i} $  were chosen such that the $ i $th intervention group had the desired proportion of observed outcomes.  The value of $ \phi_{i} $ in (\ref{miss}) represents the degree of association between the missingness and the covariate $ X $ in the $ i $th intervention group.  In this study, we assumed the same covariate effects for the probability of having a missing outcome in the two intervention groups and thus set $ \phi_{0}=\phi_{1}=1 $ in (\ref{miss}) corresponding to the OR of having  a missing outcome of 2.72 for a 1 unit change in $ X $. 

We investigated four scenarios, varying whether the baseline covariate effects on outcome and the missingness mechanisms  were the same in the two intervention groups. Table \ref{table_par} shows the parameters values used to simulate complete data (full data) and incomplete data under four different scenarios. In \textbf{scenario 1 (S1)} and \textbf{scenario 3 (S3)}, there were 30\% missing outcomes in each of the two intervention groups, while in \textbf{scenario 2 (S2)} and \textbf{scenario 4 (S4)}, there were 30\% missing outcomes in the control group and 60\% missing outcomes in the intervention group. 
\begin{table}
	\centering
	\caption{Parameters values used in the data generation process under four different scenarios: scenario 1 (\textbf{S1}), scenario 2 (\textbf{S2}), scenario 3 (\textbf{S3}) and scenario 4 (\textbf{S4})}.
	\begin{tabular}{cc|cccc}
		\toprule
		\multirow{9}{*}{\begin{turn}{90}full data \end{turn} }  	& 	 & \textbf{S1} & \textbf{S2} & \textbf{S3} & \textbf{S4} \\
		\midrule  
		& $ \beta_0 $ & \multicolumn{4}{c}{0}\\
		& $ \beta_1 $	& \multicolumn{4}{c}{1.36}\\
		& $ \beta_{2(0)} $ & 1 & 1 & 0.588 & 0.588\\
		& $ \beta_{2(1)} $ & \multicolumn{4}{c}{1}\\
		\cline{2-6}
		& $ \sigma_{x}^2 $ & \multicolumn{4}{c}{3.55} \\
		& $ \rho_{x} $   & \multicolumn{4}{c}{0.05} \\
		& $ u_{ijl} $ & \multicolumn{4}{c}{$ \mathcal{N}\left( 0,  3.37^2\right)$ }\\
		& $ \alpha_{ij} $ & \multicolumn{4}{c}{$ \mathcal{N}\left( 0,  0.18^2\right)$ }\\
		& $ \delta_{ij} $ & \multicolumn{4}{c}{$ \mathcal{N}\left( 0,  0.2^2\right)$ } \\
		& & & & &\\
		\hline
		\multirow{4}{*}{\begin{turn}{90} Missing data \end{turn} } 
		& $ \psi_0 $ & \multicolumn{4}{c}{-1.34}\\  
		& $ \psi_1 $ & -1.34 & 0.65 & -1.34 & 0.65\\
		& $ \phi_0 $ & \multicolumn{4}{c} {1} \\
		& $ \phi_1 $ & \multicolumn{4}{c} {1}  \\
		& & & & &\\
		\bottomrule
	\end{tabular}
	\label{table_par}
\end{table}
\subsection{Data analysis}
Each generated full and incomplete data sets were then analysed by $ \text{CL}_\text{U} $,  $ \text{CL}_\text{A} $, RELR and GEE. Missing outcomes were handled using CRA and MMI. We included the interaction between intervention and baseline covariate into the RELR and GEE in the case of \textbf{S3} and \textbf{S4}.  The R packages \textbf{lme4} and \textbf{geepack} were used to fit RELR and GEE, respectively. We used MMI, with a random effects logistic regression imputation model so that the imputation model  was correctly specified. For \textbf{S3} and \textbf{S4}, an interaction between intervention and baseline covariate was included in the imputation model.  The R package \texttt{jomo} \cite{matteo} was used to multiply impute each generated incomplete data set  15 times, although this package uses probit link between the mean of the outcome and the linear predictor. Both links give similar results as long as individual-level probabilities of success are not too small and not too large. The algorithm of \texttt{jomo} \cite{matteo} is basically the same with the algorithm of REALCOM-IMPUTE software for MMI \cite{carpenter2011realcom}. We used 100 burn-in iterations, which through  preliminary investigations we found to be sufficient for convergence of the posterior distribution of the parameters of our imputation model, and thinning rate 25  to avoid autocorrelation between successive draws.  When fitting the GEE models using the package \textbf{geepack} in R, we encountered convergence problems (maximum of three times out of 1000 simulation runs) in the case of \textbf{S2} and \textbf{S4}. In such situation, we fitted GEE assuming independent correlation structure.
\subsection{Simulation results}
Table \ref{table1} displays the average estimates of RD, their average estimated standard errors (SE) and coverage rates of nominal 95\% confidence intervals over 1000 simulation runs using $ \text{CL}_\text{U} $ and $ \text{CL}_\text{A} $  for each of the four scenarios.  The RD estimates using full data and using MMI followed by cluster-level analyses were unbiased for each of the four scenarios. However, CRA estimates were biased using both the $ \text{CL}_\text{U} $ and $ \text{CL}_\text{A} $ for each of the four scenarios. These results support our derived analytical results for RD estimates in Section \ref{validity_RD}. Under scenario 3, the CRA estimates of RD using both the $ \text{CL}_\text{U} $ and $ \text{CL}_\text{A} $ were coincidently close to the true value of RD.  Another simulation has been run by changing the parameter values and the estimates of RD using both the $ \text{CL}_\text{U} $ and $ \text{CL}_\text{A} $ were found to be biased (results are given in Table \ref{table_extra})in Appendix \ref{Appendix C}. As expected, the  average estimated standard errors of $ \text{CL}_\text{A} $ are smaller than that of $ \text{CL}_\text{U} $, using full data, CRA and MMI. This is because the $ \text{CL}_\text{A} $ removes the differences between the outcomes values of the two intervention groups which can be attributed to differences in the baseline covariate. MMI with adjusted DF estimates gave overcoverage for nominal 95\% confidence intervals for small number of clusters in each intervention group. 

Table \ref{table2} shows the average estimates of $ \log(\text{RR}) $, their average estimated standard errors (SE) and coverage rates for nominal 95\% confidence intervals over 1000 simulation runs using $ \text{CL}_\text{U} $ and $ \text{CL}_\text{A} $ for the all four considered scenarios. Again the full data estimates and MMI followed by cluster-level analyses estimates of $ \log(\text{RR}) $ were unbiased for all four considered scenarios. The CRA estimates  were biased using both $ \text{CL}_\text{U} $ and $ \text{CL}_\text{A} $ for all four considered scenarios. These results support our derived analytical results for RR in Section \ref{validity_RR}. MMI with adjusted DF estimates resulted overcoverage of nominal 95\% confidence intervals for small number of clusters in each intervention group. 

Recall that RELR estimates cluster-specific (also known as conditional)  intervention effect, while GEE estimates population-averaged (also known as marginal)  intervention effect.  In this study, the simulation data were generated using a RELR model  (equation (\ref{model_logitlink})), where we set $ \beta_1=1.36 $, which can be interpreted as conditional (on cluster random effects and baseline covariate $ X $) log(OR)  of developing the event of interest in the intervention group compared to the control group. For GEE, the corresponding value of $ \beta_1 $ will be smaller because the general effect of using a population averaged  model over cluster-specific model is to attenuate the regression coefficient \cite{campbell2014}. Table \ref{table3} displays the average estimates of the $ \log(\text{OR}) $, their average estimated SE and coverage rates of nominal 95\% confidence intervals using RELR and GEE. The full data estimates of GEE is slightly lower as expected than that of RELR. For GEE, the CRA and MMI estimates were compared with the mean of the full data estimates as the true population-averaged log(OR) was unknown.  The CRA estimates of RELR and GEE were unbiased with very good coverage rates. This is because we were adjusting for the baseline covariate which was associated with missingness. However, RELR  with MMI gave slightly upward biased (maximum 8.6\%) estimates of intervention effect with small number of clusters in each intervention group; while GEE with MMI gave unbiased estimates. The study by Caille \textit{et al.} \cite{Caille2014} showed similar results to ours regarding good performance of GEE with respect to bias and coverage rate using MMI. The average estimated SEs of RELR estimates using CRA were lower than that of RELR using MMI, whereas the average estimated SEs of GEE estimates using CRA and MMI are fairly similar. Therefore, there is no benefit in doing MMI over CRA when the CRA and MMI use the same set of baseline covariates.   
\begin{sidewaystable}
	\centering
	\caption{Average estimates of RD, their average estimated standard errors (SE) and coverage rates for nominal 95\% confidence intervals over 1000 simulation runs, using unadjusted cluster-level ($ \text{CL}_{\text{U}} $) and adjusted cluster-level ($ \text{CL}_{\text{A}} $) analyses with full data, CRA and MMI. Monte Carlo errors for average estimates and average estimated SEs are all less than 0.003 and 0.001, respectively. The true value of RD is 20\%.}
	\begin{tabular}{c|c|cc|cc|cc|cc|cc|cc|cc|cc|cc}
		\toprule
		&  \multirow{3}{*}{$ k $} & \multicolumn{6}{c|}{Average estimate (\%)} & \multicolumn{6}{c|}{Average estimated SE} & \multicolumn{6}{c}{Coverage (\%)} \\ 
		\cline{3-20}
		&     & \multicolumn{2}{c|}{Full} & \multicolumn{2}{c|}{CRA}    &  \multicolumn{2}{c|}{MMI}    & \multicolumn{2}{c|}{Full} & \multicolumn{2}{c|}{CRA}   &  \multicolumn{2}{c|}{MMI}    & \multicolumn{2}{c|}{Full} & \multicolumn{2}{c|}{CRA}  &  \multicolumn{2}{c}{MMI} \\
		\cline{3-20} 
		&    & $ \text{CL}_{\text{U}} $  & $ \text{CL}_{\text{A}}$   &  $ \text{CL}_{\text{U}} $  & $ \text{CL}_{\text{A}}$  &   $ \text{CL}_{\text{U}} $  & $ \text{CL}_{\text{A}}$ & $ \text{CL}_{\text{U}} $  & $ \text{CL}_{\text{A}}$   &  $ \text{CL}_{\text{U}} $  & $ \text{CL}_{\text{A}}$  &   $ \text{CL}_{\text{U}} $  & $ \text{CL}_{\text{A}}$   &  $ \text{CL}_{\text{U}} $  & $ \text{CL}_{\text{A}}$  &   $ \text{CL}_{\text{U}} $  & $ \text{CL}_{\text{A}}$   &  $ \text{CL}_{\text{U}} $  & $ \text{CL}_{\text{A}}$  \\
		\midrule
		\multirow{2}{*}{S1} 
		&  5 & 20.0 & 19.9 & 22.7 & 22.5 & 20.2 & 20.1 & 0.069 & 0.051 & 0.074 & 0.061 & 0.074 & 0.058 & 93.8 & 94.3 & 93.4 & 90.3 & 97.3 & 97.1 \\ 
		& 10 & 20.0 & 20.1 & 22.6 & 22.6 & 20.1 & 20.2 & 0.049 & 0.037 & 0.053 & 0.044 & 0.053 & 0.042 & 95.8 & 95.1 & 93.2 & 91.2 & 96.5 & 96.7 \\ 
		& 20 & 20.1 & 20.1 & 22.6 & 22.6 & 20.2 & 20.2 & 0.035 & 0.027 & 0.037 & 0.031 & 0.037 & 0.029 & 95.5 & 94.0 & 89.6 & 86.1 & 95.5 & 95.5 \\ 
		& 50 & 20.0 & 20.0 & 22.6 & 22.6 & 20.1 & 20.1 & 0.022 & 0.017 & 0.024 & 0.020 & 0.023 & 0.018 & 95.1 & 94.8 & 81.5 & 75.5 & 95.2 & 95.5 \\
		\hline
		\multirow{2}{*}{S2}
		&  5 & 20.0 & 20.0 & 11.7 & 21.9 & 19.8 & 19.8 & 0.068 & 0.052 & 0.083 & 0.070 & 0.080 & 0.066 & 95.7 & 94.8 & 86.8 & 95.4 & 98.5 & 98.8 \\ 
		& 10 & 20.2 & 20.0 & 12.0 & 21.9 & 20.1 & 19.9 & 0.049 & 0.037 & 0.059 & 0.049 & 0.056 & 0.045 & 96.1 & 95.9 & 74.4 & 94.9 & 97.5 & 97.3 \\ 
		& 20 & 19.9 & 19.9 & 11.7 & 21.9 & 20.0 & 19.9 & 0.035 & 0.027 & 0.042 & 0.036 & 0.039 & 0.032 & 95.0 & 94.5 & 52.2 & 93.0 & 94.9 & 96.2 \\
		& 50 & 20.0 & 20.1 & 11.8 & 22.0 & 20.0 & 20.1 & 0.022 & 0.017 & 0.027 & 0.023 & 0.024 & 0.020 & 95.7 & 94.9 & 13.6 & 87.5 & 95.2 & 95.7 \\
		\hline
		\multirow{2}{*}{S3}
		&  5 & 20.2 & 20.1 & 19.7 & 19.6 & 20.3 & 20.1 & 0.068 & 0.058 & 0.075 & 0.067 & 0.076 & 0.067 & 93.8 & 94.5 & 93.8 & 94.1 & 96.6 & 97.2 \\ 
		& 10 & 19.9 & 19.9 & 19.6 & 19.6 & 20.0 & 20.0 & 0.050 & 0.042 & 0.055 & 0.048 & 0.055 & 0.047 & 95.7 & 95.9 & 95.7 & 96.1 & 96.3 & 96.8 \\ 
		& 20 & 20.0 & 20.0 & 19.6 & 19.6 & 20.1 & 20.0 & 0.036 & 0.030 & 0.039 & 0.034 & 0.039 & 0.033 & 94.6 & 94.0 & 94.6 & 94.1 & 95.7 & 95.3 \\
		& 50 & 20.0 & 20.0 & 19.6 & 19.6 & 20.1 & 20.1 & 0.023 & 0.019 & 0.025 & 0.022 & 0.024 & 0.021 & 95.4 & 95.0 & 95.2 & 94.7 & 95.1 & 94.8 \\ 
		\hline
		\multirow{2}{*}{S4}
		&  5 & 20.3 & 20.2 & 9.2 & 17.4 & 20.0 & 19.9 & 0.071 & 0.058 & 0.085 & 0.076 & 0.086 & 0.075 & 94.7 & 94.0 & 82.3 & 94.4 & 98.6 & 98.8 \\ 
		& 10 & 20.1 & 20.1 & 9.2 & 17.4 & 20.2 & 20.2 & 0.050 & 0.042 & 0.060 & 0.054 & 0.059 & 0.052 & 93.9 & 94.5 & 60.9 & 92.6 & 95.9 & 96.9 \\ 
		& 20 & 19.9 & 20.0 & 8.8 & 17.1 & 19.9 & 20.0 & 0.036 & 0.030 & 0.043 & 0.038 & 0.041 & 0.037 & 95.2 & 94.1 & 29.4 & 89.5 & 95.5 & 96.2 \\ 
		& 50 & 20.0 & 20.0 & 8.8 & 17.1 & 20.0 & 20.0 & 0.023 & 0.019 & 0.027 & 0.024 & 0.026 & 0.023 & 95.0 & 95.7 &  2.3 & 80.0 & 94.8 & 94.4 \\
		\bottomrule
	\end{tabular}
	\label{table1}
\end{sidewaystable}

\begin{sidewaystable}
	\centering
	\caption{Average estimates of log(RR), their average estimated standard errors (SE) and coverage rates for nominal 95\% confidence intervals over 1000 simulation runs, using unadjusted cluster-level ($ \text{CL}_{\text{U}} $) and adjusted cluster-level ($ \text{CL}_{\text{A}} $) analyses with full data, CRA and MMI. Monte Carlo errors for average estimates and average estimated SEs are all less than 0.005 and 0.001, respectively. The true value of $\log(\text{RR})$ is 0.337.}
	\begin{tabular}{c|c|cc|cc|cc|cc|cc|cc|cc|cc|cc}
		\toprule
		&  \multirow{3}{*}{$ k $} & \multicolumn{6}{c|}{Average estimate} & \multicolumn{6}{c|}{Average estimated SE} & \multicolumn{6}{c}{Coverage (\%)} \\ 
		\cline{3-20}
		&     & \multicolumn{2}{c|}{Full} & \multicolumn{2}{c|}{CRA}    &  \multicolumn{2}{c|}{MMI}    & \multicolumn{2}{c|}{Full} & \multicolumn{2}{c|}{CRA}   &  \multicolumn{2}{c|}{MMI}    & \multicolumn{2}{c|}{Full} & \multicolumn{2}{c|}{CRA}  &  \multicolumn{2}{c}{MMI} \\
		\cline{3-20} 
		&    & $ \text{CL}_{\text{U}} $  & $ \text{CL}_{\text{A}}$   &  $ \text{CL}_{\text{U}} $  & $ \text{CL}_{\text{A}}$  &   $ \text{CL}_{\text{U}} $  & $ \text{CL}_{\text{A}}$ & $ \text{CL}_{\text{U}} $  & $ \text{CL}_{\text{A}}$   &  $ \text{CL}_{\text{U}} $  & $ \text{CL}_{\text{A}}$  &   $ \text{CL}_{\text{U}} $  & $ \text{CL}_{\text{A}}$   &  $ \text{CL}_{\text{U}} $  & $ \text{CL}_{\text{A}}$  &   $ \text{CL}_{\text{U}} $  & $ \text{CL}_{\text{A}}$   &  $ \text{CL}_{\text{U}} $  & $ \text{CL}_{\text{A}}$  \\
		\midrule
		\multirow{3}{*}{S1} 
		&  5 & 0.339 & 0.344 & 0.461 & 0.464 & 0.344 & 0.348 & 0.123 & 0.096 & 0.159 & 0.136 & 0.135 & 0.110 & 94.4 & 93.9 & 90.5 & 86.2 & 97.3 & 98.0 \\ 
		& 10 & 0.338 & 0.345 & 0.456 & 0.464 & 0.340 & 0.348 & 0.087 & 0.069 & 0.114 & 0.098 & 0.094 & 0.078 & 95.0 & 94.8 & 85.7 & 78.1 & 96.3 & 97.4 \\ 
		& 20 & 0.339 & 0.345 & 0.456 & 0.464 & 0.341 & 0.348 & 0.062 & 0.049 & 0.080 & 0.069 & 0.066 & 0.054 & 94.7 & 93.4 & 71.6 & 58.8 & 95.6 & 95.3 \\ 
		& 50 & 0.336 & 0.343 & 0.453 & 0.461 & 0.339 & 0.346 & 0.039 & 0.031 & 0.051 & 0.044 & 0.041 & 0.034 & 95.5 & 95.1 & 38.2 & 18.4 & 95.7 & 95.5 \\ 
		
		\hline
		\multirow{3}{*}{S2}
		&  5 & 0.339 & 0.346 & 0.261 & 0.515 & 0.338 & 0.344 & 0.122 & 0.096 & 0.186 & 0.161 & 0.142 & 0.119 & 95.8 & 94.7 & 94.0 & 85.8 & 98.7 & 98.8 \\ 
		& 10 & 0.341 & 0.344 & 0.266 & 0.514 & 0.340 & 0.343 & 0.087 & 0.069 & 0.130 & 0.112 & 0.098 & 0.082 & 95.7 & 95.8 & 92.2 & 69.1 & 97.5 & 97.8 \\ 
		& 20 & 0.336 & 0.342 & 0.260 & 0.512 & 0.337 & 0.343 & 0.062 & 0.049 & 0.093 & 0.081 & 0.069 & 0.057 & 95.4 & 94.6 & 88.0 & 45.4 & 95.4 & 96.2 \\ 
		& 50 & 0.337 & 0.345 & 0.263 & 0.516 & 0.337 & 0.346 & 0.039 & 0.031 & 0.059 & 0.052 & 0.043 & 0.036 & 95.7 & 94.3 & 76.2 & 6.1  & 96.7 & 95.2 \\
		\hline
		\multirow{3}{*}{S3}
		&  5 & 0.343 & 0.342 & 0.388 & 0.387 & 0.347 & 0.346 & 0.123 & 0.107 & 0.155 & 0.141 & 0.140 & 0.126 & 93.9 & 93.6 & 94.0 & 93.8 & 96.7 & 96.8 \\ 
		& 10 & 0.336 & 0.338 & 0.383 & 0.383 & 0.338 & 0.340 & 0.089 & 0.077 & 0.112 & 0.102 & 0.099 & 0.088 & 95.7 & 95.8 & 93.6 & 94.3 & 96.5 & 96.9 \\ 
		& 20 & 0.338 & 0.339 & 0.382 & 0.382 & 0.339 & 0.340 & 0.064 & 0.055 & 0.080 & 0.073 & 0.070 & 0.062 & 94.8 & 94.3 & 91.4 & 90.1 & 96.0 & 95.2 \\
		& 50 & 0.337 & 0.339 & 0.383 & 0.384 & 0.339 & 0.341 & 0.040 & 0.035 & 0.051 & 0.046 & 0.044 & 0.039 & 95.6 & 95.2 & 86.5 & 84.2 & 95.8 & 95.7\\ 
		\hline
		\multirow{3}{*}{S4}
		&  5 & 0.347 & 0.346 & 0.200 & 0.385 & 0.342 & 0.341 & 0.128 & 0.109 & 0.186 & 0.167 & 0.154 & 0.138 & 94.2 & 94.2 & 91.8 & 94.6 & 98.9 & 99.0 \\ 
		& 10 & 0.340 & 0.342 & 0.198 & 0.385 & 0.342 & 0.344 & 0.089 & 0.078 & 0.130 & 0.118 & 0.105 & 0.095 & 93.5 & 94.0 & 81.8 & 93.4 & 95.8 & 97.2 \\ 
		& 20 & 0.336 & 0.340 & 0.189 & 0.377 & 0.336 & 0.339 & 0.063 & 0.055 & 0.092 & 0.084 & 0.073 & 0.066 & 95.3 & 94.2 & 66.2 & 92.6 & 95.4 & 95.6 \\
		& 50 & 0.336 & 0.340 & 0.189 & 0.376 & 0.338 & 0.340 & 0.040 & 0.035 & 0.058 & 0.053 & 0.045 & 0.041 & 94.8 & 96.0 & 30.4 & 89.3 & 94.8 & 95.4  \\ 
		\bottomrule
	\end{tabular}
	\label{table2}
\end{sidewaystable}

\begin{sidewaystable}
	\centering
	\caption{Average estimates of log(OR), their average estimated standard errors (SE) and coverage rates for nominal 95\% confidence intervals over 1000 simulation runs, using RELR and GEE with full data, CRA and MMI. Monte Carlo errors for average estimates and average estimated SEs are all less than 0.016 and 0.003, respectively. The true value of conditional log(OR) in RELR is 1.36. The true  value of population- averaged log(OR) for GEE was empirically estimated using full data.}
	\scalebox{0.85}{
		\begin{tabular}{c|c|cc|cc|cc|cc|cc|cc|cc|cc|cc}
			\toprule
			&  \multirow{3}{*}{$ k $} & \multicolumn{6}{c|}{Average estimate} & \multicolumn{6}{c|}{Average estimated SE} & \multicolumn{6}{c}{Coverage (\%)} \\ 
			\cline{3-20}
			&     & \multicolumn{2}{c|}{Full} & \multicolumn{2}{c|}{CRA}    &  \multicolumn{2}{c|}{MMI}    & \multicolumn{2}{c|}{Full} & \multicolumn{2}{c|}{CRA}   &  \multicolumn{2}{c|}{MMI}    & \multicolumn{2}{c|}{Full} & \multicolumn{2}{c|}{CRA}  &  \multicolumn{2}{c}{MMI} \\
			\cline{3-20} 
			&    & RELR  & GEE   &  RELR  & GEE  &  RELR  & GEE & RELR  & GEE   &  RELR  & GEE  &  RELR  & GEE   &  RELR  & GEE  &   RELR  & RELR   &  RELR  & GEE  \\
			\midrule
			\multirow{2}{*}{S1}
			&  5 & 1.363 & 1.321 & 1.360 & 1.320 & 1.384 & 1.328 & 0.341 & 0.363 & 0.364 & 0.382 & 0.391 & 0.372 & 94.6 & 95.2 & 94.4 & 94.7 & 97.7 & 96.5 \\
			& 10 & 1.365 & 1.321 & 1.368 & 1.323 & 1.392 & 1.329 & 0.252 & 0.258 & 0.268 & 0.271 & 0.284 & 0.272 & 94.6 & 95.2 & 94.4 & 95.1 & 96.1 & 96.0 \\ 
			& 20 & 1.361 & 1.315 & 1.363 & 1.317 & 1.385 & 1.322 & 0.182 & 0.184 & 0.193 & 0.192 & 0.201 & 0.195 & 94.7 & 95.0 & 95.0 & 94.7 & 95.8 & 95.5 \\
			& 50 & 1.359 & 1.310 & 1.361 & 1.310 & 1.380 & 1.316 & 0.118 & 0.117 & 0.125 & 0.122 & 0.129 & 0.124 & 94.4 & 95.1 & 94.8 & 95.4 & 94.8 & 95.0 \\ 
			\hline
			\multirow{2}{*}{S2}
			&  5 & 1.345 & 1.311 & 1.368 & 1.333 & 1.402 & 1.335 & 0.336 & 0.320 & 0.405 & 0.417 & 0.456 & 0.438 & 94.7 & 94.8 & 95.5 & 94.9 & 98.6 & 98.6 \\ 
			& 10 & 1.350 & 1.309 & 1.356 & 1.313 & 1.384 & 1.308 & 0.250 & 0.258 & 0.298 & 0.301 & 0.330 & 0.317 & 93.2 & 94.4 & 94.7 & 95.4 & 97.0 & 97.1 \\ 
			& 20 & 1.358 & 1.311 & 1.352 & 1.305 & 1.376 & 1.301 & 0.184 & 0.185 & 0.215 & 0.213 & 0.232 & 0.224 & 94.8 & 95.8 & 95.0 & 94.9 & 96.7 & 96.4 \\ 
			& 50 & 1.366 & 1.316 & 1.367 & 1.318 & 1.389 & 1.316 & 0.118 & 0.117 & 0.138 & 0.135 & 0.146 & 0.141 & 95.3 & 95.7 & 95.0 & 95.0 & 95.8 & 96.0 \\ 
			\hline
			\multirow{2}{*}{S3} 
			&  5 & 1.391 & 1.353 & 1.407 & 1.367 & 1.434 & 1.374 & 0.343 & 0.358 & 0.392 & 0.400 & 0.414 & 0.389 & 94.8 & 94.1 & 95.2 & 94.4 & 97.7 & 97.4 \\ 
			& 10 & 1.352 & 1.307 & 1.359 & 1.314 & 1.385 & 1.320 & 0.254 & 0.259 & 0.284 & 0.286 & 0.299 & 0.285 & 92.8 & 94.1 & 94.0 & 94.5 & 95.4 & 95.0 \\ 
			& 20 & 1.372 & 1.326 & 1.370 & 1.325 & 1.395 & 1.330 & 0.183 & 0.184 & 0.204 & 0.202 & 0.212 & 0.203 & 93.2 & 94.4 & 93.2 & 94.1 & 94.1 & 94.1 \\
			& 50 & 1.363 & 1.313 & 1.363 & 1.313 & 1.386 & 1.317 & 0.118 & 0.117 & 0.132 & 0.127 & 0.135 & 0.129 & 95.1 & 95.1 & 94.8 & 95.5 & 95.4 & 95.4 \\
			\hline
			\multirow{2}{*}{S4}
			&  5 & 1.375 & 1.336 & 1.413 & 1.378 & 1.476 & 1.390 & 0.346 & 0.366 & 0.497 & 0.493 & 0.535 & 0.505 & 94.5 & 95.2 & 97.0 & 94.0 & 98.6 & 98.5 \\ 
			& 10 & 1.366 & 1.325 & 1.377 & 1.334 & 1.431 & 1.342 & 0.252 & 0.258 & 0.353 & 0.351 & 0.375 & 0.357 & 94.6 & 95.3 & 95.3 & 94.6 & 96.5 & 96.6 \\ 
			& 20 & 1.376 & 1.328 & 1.387 & 1.339 & 1.432 & 1.346 & 0.183 & 0.184 & 0.252 & 0.247 & 0.266 & 0.251 & 94.7 & 94.8 & 94.3 & 94.4 & 94.5 & 94.8 \\ 
			& 50 & 1.360 & 1.312 & 1.362 & 1.313 & 1.397 & 1.317 & 0.118 & 0.117 & 0.160 & 0.156 & 0.167 & 0.157 & 95.4 & 95.7 & 94.8 & 94.5 & 94.4 & 94.2 \\
			\bottomrule
		\end{tabular}}
		\label{table3}
	\end{sidewaystable}
	
	\section{Example}
	\label{example}
	We now illustrate the methods compared here using the data  from a factorial cluster randomised trial designed to investigate the impact of two interventions among school children in class 1 and class 5 on the south coast of Kenya \cite{Halliday2014}. The interventions were  intermittent screening and treatment (IST) for malaria on the health and education of school children in class 1 and class 5;  and  a literacy intervention (LIT) on education only being applied in class 1. One hundred and one government primary schools were randomised to one of the four groups receiving either (i) IST alone (25 schools); (ii) LIT alone (25 schools); (iii) both IST and LIT (26 schools); or (iv) neither IST nor LIT (25 schools). On average, the number of  children per school in the four groups  were, respectively,  107 ( standard deviation (SD)=7.54 ), 99 (SD=17.84), 103 (SD=6.28) and 102 (SD=7.51). The primary outcomes were anaemia at either 12 or 24 months and educational achievement at 9 months and 24 months assessed by a battery of tests of reading, writing and arithmetic. Baseline characteristics of the school (school mean exam score and school size), the child (age, sex,  sleep under net and baseline anaemia) and the household (paternal education and  household size) were collected.  For the purpose of illustration, we restricted attention to anaemia (binary) measured at the 24 months follow-up. A paper published  based on this study \cite{Halliday2014} showed  no evidence of interaction between the two interventions in class 1 where both were implemented. We therefore merged the groups (i) and (iii) where IST was implemented and considered this as the  intervention group;  and merged the groups (ii) and (iv) where IST was not implemented and considered this as the control group. The control group and the intervention group consisted of  2502 and 2674 children, respectively; and among them 475 (18.98\%) and 501 (18.74\%) had missing anaemia at 24 months, respectively. The covariate baseline anaemia had some missing values as well. To illustrates our methods for the case where only outcomes are missing and all baseline covariates are fully observed, we excluded the children from the analysis with missing baseline anaemia value. Hence, in our analysis, the control group and the intervention group consisted of  2373 and 2451 children, respectively; and among them 430 (18.12\%) and 424 (17.30\%) had missing anaemia at 24 months, respectively.
	
	The original trial's prespecified analysis planned to adjust for the baseline covariates age, sex, exam score, literacy group and baseline anaemia.  In our analysis, first we investigated the association of the baseline covariates (age, sex, exam score, literacy group and baseline anaemia) with anaemia at 24 months and with the probability of anaemia at 24 months being missing by fitting random effects logistic regression models.  Table \ref{datatable1} displays the estimates of conditional log odds ratios of the two models. Age and baseline anaemia were strongly associated with anaemia at 24 months and there was no  evidence of interaction between IST intervention and baseline covariates in the model for anaemia at 24 months. Older children were more likely to have anaemia at 24 months missing; and children receiving literacy intervention were less likely to have anaemia at 24 months missing. There was weak evidence of interaction between IST intervention and literacy group on the missingness of anaemia at 24 months. Based on these analyses a working assumption is that missingness of anaemia at 24 months depends mainly on age, and that this dependence does not differ between the two intervention groups as there was no evidence of interaction between IST intervention and age.  
	\begin{table}[H]
		\centering
		\caption{Estimates of log odds ratios as measures of association of the baseline covariates with anaemia at 24 months and with the probability of anaemia at 24 months being missing}
		\begin{threeparttable}
			\centering
			\begin{tabular}{lccc|ccc}
				\toprule
				& \multicolumn{3}{c|}{Anaemia} & \multicolumn{3}{c}{Missingness of anaemia}\\
				\midrule
				& Estimate & Std. Error & p-\text{value} & Estimate & Std. Error & p-\text{value} \\ 
				\midrule
				Intercept                         & -1.72 & 0.81 &        0.03  & -2.10 & 0.60 & 0.00  \\ 
				IST (intervention)                &  0.36 & 1.10 &        0.74  & -0.27 & 0.83 & 0.74  \\ 
				Age (years)                       &  0.07 & 0.02 & $ <0.001 $ &  0.06 & 0.02 & $ <0.001 $ \\ 
				Sex (male vs female)              & -0.04 & 0.10 & 0.73         & -0.08 & 0.11 & 0.48  \\ 
				Exam score                        &  0.00 & 0.00 & 0.77         &  0.00 & 0.00 & 0.91  \\ 
				Literacy group                    &  0.06 & 0.19 & 0.74         & -0.28 & 0.13 & $ 0.03 $  \\ 
				Baseline anaemia                  &  1.57 & 0.11 & $ <0.001 $ &  0.09 & 0.11 & 0.42  \\ 
				IST: Age \tnote{*}                &  0.01 & 0.03 & 0.62         &  0.04 & 0.03 & 0.12  \\ 
				IST: Sex \tnote{*}                &  0.10 & 0.14 & 0.49         & -0.18 & 0.15 & 0.24  \\ 
				IST: Exam score \tnote{*}         &  0.00 & 0.00 & 0.59         &  0.00 & 0.00 & 0.62  \\ 
				IST: Literacy group \tnote{*}     &  0.37 & 0.26 & 0.15         &  0.38 & 0.19 & $ 0.04 $  \\ 
				IST: Baseline anaemia \tnote{*}   & -0.19 & 0.15 & 0.19         & -0.03 & 0.15 & 0.86  \\ 
				\bottomrule
			\end{tabular}
			\begin{tablenotes}
				\item[*] Interaction terms
			\end{tablenotes}
		\end{threeparttable}
		\label{datatable1}
	\end{table}
	
	\begin{sidewaystable}
		\caption{Risk difference, risk ratio and odds ratio estimates using CRA and MMI for the IST intervention trial data. }
		\begin{threeparttable}
			\centering
			\begin{tabular}{lcccccc}
				\toprule
				Analysis Approach & & $ N_0 $ & $ N_1 $ & Risk difference (RD) & Risk ratio (RR) & Odds ratio (OR)\\
				\hline
				&    &   &  & Estimate (95\% CI) & Estimate (95\% CI) & Estimate ( 95\% CI )\\
				\hline\hline
				\multicolumn{7}{l}{\textbf{Cluster-level Analysis}\tnote{a}}\\
				\multicolumn{7}{l}{\textbf{}}\\
				\multicolumn{7}{l}{\textbf{CRA}}\\
				Unadjusted & & 2027 & 2173 & 0.019 (-0.040,  0.077) & 1.047 (0.908, 1.208)  &   \\
				Adjusted   & & 1935 & 2027 & 0.022 (-0.033,  0.077) & 1.037 (0.908, 1.185) &   \\
				
				& & & &                        &                        &  \\
				\multicolumn{7}{l}{\textbf{MMI}}\\
				Unadjusted & & 2373 & 2451  & 0.021 (-0.038, 0.080) & 1.053 (0.911, 1.218)  &  \\
				Adjusted   & & 2373 & 2451  & 0.017 (-0.035, 0.070) & 1.040 (0.910, 1.189)  &   \\ 
				
				\midrule
				\multicolumn{7}{l}{\textbf{Individual-level Analysis}}\\ 
				\multicolumn{7}{l}{\textbf{}}\\
				\multicolumn{7}{l}{\textbf{CRA}}\\
				\multicolumn{7}{l}{\textbf{RELR}}\\
				Unadjusted  & & 2027 & 2173 & & -  & 1.090 (0.841, 1.414)\\
				Adjusted    & & 1935 & 2027 & & -  & 1.088 (0.839, 1.409)\\
				
				\multicolumn{7}{l}{\textbf{GEE}\tnote{b}}\\
				Unadjusted   & & 2027 & 2173 & & 1.048 (0.908, 1.209) & 1.082 (0.850, 1.378)\\
				Adjusted     & & 1935 & 2027 & & 1.019 (0.911, 1.141) & 1.070 (0.842, 1.359)\\
				& & & & & &  \\
				\multicolumn{7}{l}{\textbf{MMI}}\\
				\multicolumn{7}{l}{\textbf{RELR}}\\
				Unadjusted  & & 2373 & 2451   & & - & 1.101 (0.849, 1.428)\\
				Adjusted    & & 2373 & 2451   & & - & 1.089 (0.841, 1.413)\\
				\multicolumn{7}{l}{\textbf{GEE}}\\
				Unadjusted   & & 2373 & 2451   & & 1.053 (0.912, 1.215) & 1.090 (0.856, 1.389) \\
				Adjusted     & & 2373 & 2451   & & 1.019 (0.911, 1.140) & 1.072 (0.843, 1.363) \\
				\bottomrule
			\end{tabular}
			\begin{tablenotes}
				\item[a] Cluster-level analysis was used to estimate the risk difference and the risk ratio.
				\item[b] GEE was used to estimate the risk ratio using log link and to estimate the marginal odds ratio using logit link.  
			\end{tablenotes}
		\end{threeparttable}
		\label{datatable2}
	\end{sidewaystable}
	
	We analysed the data using the methods $ \text{CL}_{\text{U}} $, $ \text{CL}_{\text{A}} $, RELR and GEE; assuming that the missingness in anaemia at 24 months depends on the baseline covariates age, but conditioning on age, not on the anaemia at 24 months itself, i.e. a CDM mechanism. GEE models were fitted assuming both logit and log links for the true outcome model to estimate odds ratio and risk ratio, respectively. The missing anaemia at 24 months were handled using CRA and MMI. The RELR, GEE and adjusted cluster-level analysis were adjusted for the baseline covariates age,  sex, school mean exam score, literacy group and baseline anaemia.  MMI was done using the R package \texttt{jomo} \cite{matteo}, with an imputation model adjusted for the aforementioned baseline covariates. We used 100 imputed datasets in MMI. GEE with log link after MMI was not congenial with the imputation model, as the imputation model used probit link. The estimates and confidence intervals (CIs) of RD, RR and OR obtained by CRA and MMI are displayed in Table \ref{datatable2}. The column $ N_0 $ and $ N_1 $ in Table \ref{datatable2}  represent the number of children in the control and intervention groups, respectively. All measures showed no evidence of IST intervention effect in improving health of school children by alleviating anaemia. The CRA  estimates of RD and RR using cluster-level analyses are very similar to the corresponding estimates obtained by MMI. This is because CRA is valid in this case as there is no evidence of intervention effect and no evidence of interaction between covariates and intervention. The estimates and CIs of unadjusted and adjusted OR obtained by CRA were found to be very close to the corresponding estimates obtained by MMI. This is because, as we found in our simulation results, there is no gain in terms of bias or efficiency of the estimates using MMI over CRA  as long as  the same set of predictors of missingness are used by both methods.   
	\vskip 2cm
	\section{Discussion and conclusion}
	\label{dis_con}
	In this paper, we showed analytically and through simulations that cluster-level analyses for estimating RD using complete records are valid only when there is no intervention effect in truth and the intervention groups have the same missingness mechanism and the same covariate effect in the outcome model.  For estimating RR, cluster-level analyses using complete records are valid if the true data generating model has log link and the intervention groups have the same missingness mechanism and the same covariate effect in the outcome model. However, if the true data generating model has logit link, cluster-level analyses using complete records for estimating RR are valid only  when there is no intervention effect in truth and the intervention groups have the same missingness mechanism and the same covariate effect in the outcome model. But, in practice, it is impossible to know in advance whether there is an  intervention effect. We therefore caution researchers that cluster-level analyses using complete records, assuming logit link for the true data generating model,  in general results in biased inferences in CRTs. However, when the true data generating model follows a log link and the parameter of interest is RR, cluster-level analyses using complete records give valid inferences if the intervention groups have the same missingness mechanism and the same covariates effect in the outcome model.
	
	In contrast, MMI followed by cluster-level analyses gave unbiased estimates of RD and RR regardless of whether missingness mechanisms were the same or different between the intervention groups and whether there is an interaction between intervention and baseline covariate in the outcome model, provided that an interaction was allowed for in the imputation model when required. However, MMI resulted in overcoverage for the nominal 95\% confidence interval with small number of clusters in each intervention group. Similar results were found for continuous outcomes in CRTs by Hossain \textit{et al.} \cite{Hossain2016}.
	
	The full data estimates of conditional  $ \log(\text{OR}) $ using  RELR were unbiased with good coverage rates. These results differ from the results found by Ma \textit{et al.} \cite{Ma2013}, where they concluded that full data estimates using RELR were biased. As noted previously, we believe this is because they generated the data in such a way that they knew  what the true population-averaged $ \log(\text{OR}) $ was, but after fitting RELR, they compared the estimates of conditional (on the cluster random effects) $ \log(\text{OR}) $ with the true population averaged  $ \log(\text{OR}) $. As noted earlier, population averaged $ \log(\text{OR}) $ is marginal with respect to the cluster random effects \cite{Faraway2006}.
	
	The CRA estimates of conditional $ \log(\text{OR}) $ using  RELR were unbiased with coverage rates close to the nominal level regardless of whether the missingness mechanism is the same or different between the intervention groups and whether there is an interaction between the intervention and baseline covariate in the data generating model for outcome, provided that if there is an interaction in the data generating model for the outcome then this interaction is included in the model fitted to the data. This conclusion contradicts the results of a previous study by Ma \textit{et al.} \cite{Ma2013}, where they found that CRA estimates using RELR are biased under covariate dependent missingness (CDM) assumption. Again we believe this is because they compared RELR estimates of the conditional $ \log(\text{OR}) $ to the true marginal $ \log(\text{OR}) $. The conclusions of Ma \textit{et al.} \cite{Ma2013} have subsequently been cited in a recent textbook on CRT design and analysis \cite{campbell2014}. We hope that our results and explanations help in understanding some of the surprising results and conclusion in Ma \textit{et al.} \cite{Ma2011, Ma2012comparing, Ma2013}. In our study, we also found that the RELR with MMI gave slightly upward biased estimates of  conditional $ \log(\text{OR}) $ for small number of clusters in each intervention groups.  
	
	GEE using CRA and MMI gave unbiased estimates of population averaged $ \log(\text{OR}) $ with coverage rates close to the nominal level regardless of whether the missingness mechanism was the same between the intervention groups and whether there was an interaction between the intervention group and baseline covariate in the data generating model. Similar results had been found by Ma \textit{et al.} \cite{Ma2013} for GEE in terms of bias, although as described earlier, in their data generating mechanism the covariate was generated independently of the outcome.
	
	In this study, we have assumed baseline covariate dependent missingness assumption for binary outcome, which is an example of MAR as our baseline covariate was fully observed. In practice, it cannot be identified on the basis of the observed data  which missingness assumption is appropriate \cite{White2010,carpenterKen2013}. Therefore, sensitivity analyses should be performed \cite[Ch. 10]{carpenterKen2013} to explore whether  inferences are robust to the primary working assumption regarding the missingness mechanism. Furthermore, we focused on studies with only one individual-level baseline covariate; the methods described can be extended to more than one baseline covariate.
	
	In conclusion, as long as both MMI and CRA use the same set of baseline covariates, RELR or GEE using complete records can be recommended as the primary analysis approach for CRTs with missing binary outcomes if we are willing to assume that the missingness depends on baseline covariates and conditional on these, not on the outcome.  In addition, where the aim is to estimate RD or RR,  MMI can be used followed by cluster-level analysis to get valid estimates under the covariate dependent missingness assumption for missing binary outcomes, but one should be cautious when making inferences as this approach results in overcoverage for small number of clusters in each intervention group.
	
	\section*{Acknowledgements} 
	A Hossain was supported by the Economic and Social Research Council (ESRC), UK, via Bloomsbury Doctoral Training Centre (ES/J5000021/1). K Diaz-Ordaz was funded by Medical Research Council (MRC) career development award in Biostatistics (MR/L011964/1). J W Bartlett's contribution to this paper was partly supported by MRC fellowship (MR/K02180X/1) while he was a member of the Department of Medical Statistics, London School of Hygiene \& Tropical Medicine (LSHTM). The authors would like to thank Professor Elizabeth Allen, Department of Medical Statistics, LSHTM, for giving us permission to use the IST intervention trial data. We also would like to thank all teachers, children and parents who participated in this trial.

	
	\begin{appendices}
		\section{Unbiasedness  of $ \widehat{\text{RD}}_{\text{unadj}} $  and consistency of $ \widehat{\text{RR}}_{\text{unadj}} $ with full data}
		\label{app1}
		\begin{eqnarray}
		\text{E}\left(  \bar{p}_i\right)  &=& \frac{1}{k}\text{E}\left( p_{ij}\right) \nonumber\\
		&=& \frac{1}{mk}\sum_{j=1}^{k} \sum_{l=1}^{m} \text{E}\left( Y_{ijl}\right)\nonumber \\
		&=& \pi_i \nonumber
		\end{eqnarray}	
		where $ \pi_i $ is the true proportion of success in the $ i $th intervention group. Then
		\begin{eqnarray}
		\text{E}\left(  \widehat{\text{RD}}_{\text{unadj}}\right)  &=& \text{E}\left( \bar{p}_1 - \bar{p}_1\right) \nonumber\\
		&=& \pi_1-\pi_0\nonumber\\
		&=& \text{RD}\nonumber
		\end{eqnarray}	
		\begin{eqnarray}
		\text{E}\left( \widehat{\text{RR}}_{\text{unadj}} \right) = \text{E}\left( \frac{\bar{p}_1}{\bar{p}_0} \right) & \longrightarrow & \frac{\text{E}\left( \bar{p_1} \right) }{\text{E}\left( \bar{p_0} \right)}\text{~as~} k\to \infty\nonumber\\
		&=& \frac{\pi_1}{\pi_0}\nonumber\\
		&=& \text{RR} \nonumber
		\end{eqnarray}
		\section{Derivation of equation (\ref{equation_RRadj})}
		\label{Appendix B}
		\begin{eqnarray}
		\text{E}\left( \widehat{\text{RR}}_{\text{adj}}^{\text{cr}}\right) &\longrightarrow &\frac{\text{E}\left( \bar{\epsilon}_1^{\,r\text{(cr)}}\right) }{\text{E}\left( \bar{\epsilon}_0^{\,r\text{(cr)}}\right) } \text{ as } k\to \infty \nonumber\\
		&=& \frac{\text{E}\left(\frac{1}{k}\sum_{j=1}^{k}\frac{n_{1j}^{\text{cr}}}{\hat{n}_{1j}^{\text{cr}}} \right) }{\text{E}\left(\frac{1}{k}\sum_{j=1}^{k}\frac{n_{0j}^{\text{cr}}}{\hat{n}_{0j}^{\text{cr}}} \right) }\nonumber\\
		&=& \frac{\text{E}\left(\frac{n_{1j}^{\text{cr}}}{\hat{n}_{1j}^{\text{cr}}} \right) }{\text{E}\left(\frac{n_{0j}^{\text{cr}}}{\hat{n}_{0j}^{\text{cr}}} \right) }\nonumber\\
		&=& \frac{\text{E}\left(\frac{p_{1j}^{\text{cr}}}{\hat{p}_{1j}^{\text{cr}}} \right) }{\text{E}\left(\frac{p_{0j}^{\text{cr}}}{\hat{p}_{0j}^{\text{cr}}} \right) }\nonumber\\
		&\longrightarrow & \frac{\text{E}\left(p_{1j}^{\text{cr}} \right) }{\text{E}\left(p_{0j}^{\text{cr}} \right)}\times 
		\frac{\text{E}\left(\hat{p}_{0j}^{\text{cr}} \right) }{\text{E}\left(\hat{p}_{1j}^{\text{cr}} \right)}\text{ as } m\to \infty\nonumber\\
		&=& \frac{\text{E}\left(\bar{p}_{1}^{\text{cr}} \right) }{\text{E}\left(\bar{p}_{0}^{\text{cr}} \right)}\times 
		\frac{\text{E}\left(\hat{p}_{0j}^{\text{cr}} \right) }{\text{E}\left(\hat{p}_{1j}^{\text{cr}} \right)}\nonumber\\
		&\longrightarrow & \text{E} \left( \widehat{\text{RR}}_{\text{unadj}}^{\text{cr}} \right) \frac{\text{E}\left(\hat{p}_{0j}^{\text{cr}} \right) }{\text{E}\left(\hat{p}_{1j}^{\text{cr}} \right)} \text{ as } (k,m)\to \infty. \nonumber 
		\end{eqnarray}

		\section{Additional simulation results for RD}
		\label{Appendix C}
		\setcounter{table}{0}
		\renewcommand{\thetable}{A\arabic{table}}
		
		\begin{sidewaystable}
			\centering
			\caption{Average estimates of RD, their average estimated standard errors (SE) and coverage rates for nominal 95\% confidence intervals over 1000 simulation runs, using unadjusted cluster-level ($ \text{CL}_{\text{U}} $) and adjusted cluster-level ($ \text{CL}_{\text{A}} $) analyses with full data, CRA and MMI. The true value of RD is 15\%.}
			\begin{tabular}{c|c|cc|cc|cc|cc|cc|cc|cc|cc|cc}
					\toprule
					&  \multirow{3}{*}{$ k $} & \multicolumn{6}{c|}{Average estimate (\%)} & \multicolumn{6}{c|}{Average estimated SE} & \multicolumn{6}{c}{Coverage (\%)} \\ 
					\cline{3-20}
					&     & \multicolumn{2}{c|}{Full} & \multicolumn{2}{c|}{CRA}    &  \multicolumn{2}{c|}{MMI}    & \multicolumn{2}{c|}{Full} & \multicolumn{2}{c|}{CRA}   &  \multicolumn{2}{c|}{MMI}    & \multicolumn{2}{c|}{Full} & \multicolumn{2}{c|}{CRA}  &  \multicolumn{2}{c}{MMI} \\
					\cline{3-20} 
					&    & $ \text{CL}_{\text{U}} $  & $ \text{CL}_{\text{A}}$   &  $ \text{CL}_{\text{U}} $  & $ \text{CL}_{\text{A}}$  &   $ \text{CL}_{\text{U}} $  & $ \text{CL}_{\text{A}}$ & $ \text{CL}_{\text{U}} $  & $ \text{CL}_{\text{A}}$   &  $ \text{CL}_{\text{U}} $  & $ \text{CL}_{\text{A}}$  &   $ \text{CL}_{\text{U}} $  & $ \text{CL}_{\text{A}}$   &  $ \text{CL}_{\text{U}} $  & $ \text{CL}_{\text{A}}$  &   $ \text{CL}_{\text{U}} $  & $ \text{CL}_{\text{A}}$   &  $ \text{CL}_{\text{U}} $  & $ \text{CL}_{\text{A}}$  \\
					\midrule
					\multirow{2}{*}{S1} 
					& 5  & 14.9 & 14.9 & 16.7 & 16.6 & 15.0 & 15.0 & 0.071 & 0.053 & 0.075 & 0.063 & 0.076 & 0.060 & 94.5 & 95.9 & 94.5 & 95.2 & 96.6 & 98.7  \\ 
					& 10 & 15.1 & 15.1 & 16.9 & 16.8 & 15.2 & 15.2 & 0.050 & 0.038 & 0.054 & 0.045 & 0.054 & 0.042 & 94.2 & 93.6 & 93.3 & 92.3 & 95.1 & 94.9 \\ 
					& 20 & 15.1 & 15.0 & 16.8 & 16.7 & 15.2 & 15.1 & 0.036 & 0.027 & 0.038 & 0.032 & 0.038 & 0.030 & 94.5 & 94.4 & 92.8 & 90.8 & 95.3 & 94.5 \\ 
					& 50 & 15.1 & 15.1 & 16.7 & 16.7 & 15.0 & 15.0 & 0.023 & 0.017 & 0.024 & 0.020 & 0.023 & 0.018 & 94.6 & 95.3 & 88.9 & 85.2 & 90.4 & 90.7 \\ 
					\hline
					\multirow{2}{*}{S2}
					& 5  & 14.9 & 14.9 & 5.3  & 15.9 & 14.9 & 15.1 & 0.070 & 0.052 & 0.082 & 0.069 & 0.083 & 0.068 & 94.1 & 95.2 & 82.0 & 95.6 & 97.6 & 98.8 \\ 
					& 10 & 14.9 & 15.1 & 5.5  & 16.0 & 15.1 & 15.0 & 0.050 & 0.038 & 0.059 & 0.050 & 0.058 & 0.048 & 94.9 & 95.1 & 68.6 & 95.1 & 95.5 & 97.0 \\ 
					& 20 & 15.1 & 15.1 & 5.5  & 16.0 & 15.0 & 14.9 & 0.036 & 0.027 & 0.042 & 0.036 & 0.041 & 0.033 & 94.8 & 94.5 & 40.3 & 94.5 & 95.4 & 95.8 \\ 
					& 50 & 15.0 & 15.0 & 5.5  & 16.0 & 15.0 & 15.0 & 0.023 & 0.017 & 0.027 & 0.023 & 0.025 & 0.021 & 94.6 & 94.3 & 6.0  & 94.2 & 95.8 & 96.8 \\ 	
					\hline
					\multirow{2}{*}{S3}
					&  5 & 15.2 & 15.2 & 13.2 & 13.2 & 15.5 & 15.4 & 0.072 & 0.061 & 0.078 & 0.070 & 0.081 & 0.071 & 95.6 & 96.4 & 93.7 & 94.1 & 97.6 & 99.1 \\ 
					& 10 & 15.0 & 15.0 & 12.9 & 12.9 & 15.0 & 15.1 & 0.052 & 0.044 & 0.056 & 0.050 & 0.057 & 0.050 & 94.8 & 94.8 & 93.5 & 93.8 & 96.8 & 96.8 \\ 
					& 20 & 15.0 & 15.0 & 13.0 & 12.9 & 15.1 & 15.1 & 0.036 & 0.031 & 0.039 & 0.035 & 0.040 & 0.035 & 94.3 & 93.9 & 92.7 & 90.0 & 95.6 & 95.4 \\ 
					& 50 & 15.1 & 15.2 & 13.0 & 13.0 & 15.1 & 15.2 & 0.023 & 0.020 & 0.025 & 0.023 & 0.025 & 0.022 & 94.7 & 96.2 & 85.4 & 86.5 & 94.6 & 95.4 \\
					\hline
					\multirow{2}{*}{S4}
					&  5 & 15.1 & 14.9 & 1.8 & 9.5 & 15.0 & 14.8 & 0.072 & 0.061 & 0.084 & 0.076 & 0.089 & 0.080 & 96.0 & 95.5 & 73.8 & 92.8 & 98.7 & 99.0 \\ 
					& 10 & 15.1 & 15.1 & 1.9 & 9.8 & 15.1 & 15.0 & 0.051 & 0.044 & 0.061 & 0.055 & 0.062 & 0.056 & 93.6 & 94.0 & 45.6 & 86.3 & 96.6 & 97.1 \\ 
					& 20 & 15.1 & 15.0 & 1.7 & 9.7 & 15.1 & 15.0 & 0.036 & 0.031 & 0.043 & 0.039 & 0.043 & 0.039 & 94.4 & 96.0 & 15.8 & 72.2 & 96.1 & 97.0 \\ 
					& 50 & 15.0 & 15.0 & 1.8 & 9.8 & 15.1 & 15.1 & 0.023 & 0.020 & 0.027 & 0.025 & 0.027 & 0.024 & 94.6 & 94.4 & 0.3 & 42.6 & 95.2 & 95.1 \\ 
					\bottomrule
				\end{tabular}
				\label{table_extra}
			\end{sidewaystable}

	\end{appendices}

\end{document}